\documentclass[12pt]{article}
\usepackage{graphicx}
\renewcommand{\baselinestretch}{.98}
\begin{document}

\begin{center}
{\Large \bf Event patterns (particle scatter plots) extracted from
charged particle spectra in $pp$ and Pb-Pb collisions at 2.76 TeV}

\vskip0.75cm

Ya-Hui~Chen$^{a}$, Fu-Hu~Liu$^{a,}${\footnote{E-mail:
fuhuliu@163.com; fuhuliu@sxu.edu.cn}}, Sakina~Fakhraddin$^{b,c}$,
Magda~A.~Rahim$^{b,c}$, Mai-Ying~Duan$^a$

\vskip0.25cm

{\small

\it $^a$Institute of Theoretical Physics, Shanxi University,
Taiyuan, Shanxi 030006, China

$^b$Department of Physics, College of Science and Arts at
Riyadh~Al~Khabrah, Qassim~University, Buraidah 51452, Al~Qassim,
Saudi~Arabia

$^c$Department of Physics, Faculty of Science, Sana'a University,
P.O. Box 1247, Sana'a, Yemen}

\end{center}

\vskip0.5cm

{\bf Abstract:} The transverse momentum ($p_T$) and pseudorapidity
($\eta$) spectra of charged particles produced in proton-proton
($pp$) and lead-lead (Pb-Pb) collisions at the large hadron
collider (LHC) are described by a hybrid model. In the model, the
$p_T$ spectrum is described by a two-component distribution which
contains an inverse power-law suggested by the QCD (Quantum
Chromodynamic) calculus and an Erlang distribution resulted from a
multisource thermal model. The $\eta$ spectrum is described by a
Gaussian rapidity ($y$) distribution resulted from the Landau
hydrodynamic model and the two-component $p_T$ distribution, where
the conversion between $y$ and $\eta$ is accurately considered.
The modelling results are in agreement with the experimental data
measured by the ATLAS Collaboration in $pp$ collisions at
center-of-mass energy $\sqrt{s}=2.76$ TeV and in Pb-Pb collisions
at center-of-mass energy per nucleon pair $\sqrt{s_{NN}}=2.76$
TeV. Based on the parameter values extracted from $p_T$ and $\eta$
or $y$ spectra, the event patterns or particle scatter plots in
three-dimensional velocity and momentum spaces are obtained.
\\

\\


\vskip1.0cm

{\section{Introduction}}

The relativistic heavy ion collider (RHIC) and the large hadron
collider (LHC) have been opening a new epoch for high-energy
nucleus-nucleus (heavy ion) collisions, in which not only the
quark-gluon plasma (QGP) is created, but also more abundant
phenomena on multi-particle productions are discovered [1--7].
High-energy nucleus-nucleus collisions at the RHIC and LHC can
form a condition of high temperature and density. The evolution
and decay of the interacting system at high temperature and
density is a complex process, in which only limited information
can be measured in experiments due to technical and economical
reasons. To understand the whole interacting system as minutely as
possible, the method of event reconstruction and reappearance is
used in the modelling analyses. By using the method of event
reconstruction and reappearance, we can obtain partly the event
patterns or particle scatter plots at the stage of kinetic
freeze-out of the interacting system. Generally, the event
patterns (particle scatter plots) are expected to be different in
descriptions for different particles produced in different
collisions at different energies.

To reconstruct and reappear the event patterns (particle scatter
plots) at kinetic freeze-out, we need at least the transverse
momentum ($p_T$) and rapidity ($y$) or pseudorapidity ($\eta$)
spectra. The $p_T$ spectrum reflects the transverse excitation
degree, and the $y$ or $\eta$ spectrum reflects the longitudinal
expansion degree, of the interacting system. To describe the $p_T$
spectrum, one has used more than ten functions which include, but
are not limited to, the standard distribution [8, 9], Tsallis
statistics [10--12], Schwinger mechanism [13--16], Erlang
distribution [17], inverse power-law [18--20], and so forth. Among
these functions, some of them (standard distribution, Tsallis
statistics, and Erlang distribution) are based on thermal or
statistical reason, and others (Schwinger mechanism and inverse
power-law) are based on the QCD (Quantum Chromodynamic) calculus.
Generally, the spectrum in low-$p_T$ region is described by the
thermal and statistical distributions, and the spectrum in
high-$p_T$ region is described by the QCD calculus. Exceptionally,
the Schwinger mechanism describes only the spectrum in low-$p_T$
region, and the Tsallis statistics describes a wide spectrum. In
most case, one needs a two-component distribution to describe the
$p_T$ spectrum.

To describe the $y$ spectrum, one has the Gaussian distribution
[21--24], two-Gaussian distribution [25], three-Gaussian
distributions [26], and other modelling descriptions such as the
three-fireball model [27--32], the three-source relativistic
diffusion model [33--36], and the model with two Tsallis (or
Boltzmann-Gibbs) clusters of fireballs [37--39]. The Gaussian $y$
distribution is resulted from the Landau hydrodynamic model and
its revisions [21--24], the two-Gaussian $y$ distribution [25] can
be resulted from a two-component Landau hydrodynamic model in
which the two Gaussian distributions correspond to the
contributions in the backward and forward $y$ regions,
respectively, and the three-Gaussian $y$ distribution is resulted
from the three-component Landau hydrodynamic model in which the
third Gaussian distributions correspond to the contribution in the
central $y$ region [26]. It should be noticed that the backward
and forward $y$ regions are relative in collider experiments. Even
if for the backward, forward, and central $y$ regions, there are
alternative methods to describe the $y$ spectrum such as the
three-fireball [27--32] or three-source model [33--36] which
results in other $y$ distributions. Most models describe the $y$
spectrum to be arithmetic solutions than analytic one.

In our recent works [40, 41], the event patterns (particle scatter
plots) extracted from the spectra of net-baryons produced in
central gold-gold (Au-Au) collisions at RHIC energies, and from
the spectra of $Z$ bosons and quarkonium states (some charmonium
$c\bar c$ mesons and bottomonium $b\bar b$ mesons) produced in
proton-proton ($pp$) and lead-lead (Pb-Pb) collisions at LHC
energies, were reported. As a successor, the present work presents
the event patterns (particle scatter plots) extracted from the
spectra of charged particles produced in $pp$ collisions at the
center-of-mass energy $\sqrt{s}=2.76$ TeV and in Pb-Pb collisions
at the center-of-mass energy per nucleon pair $\sqrt{s_{NN}}=2.76$
TeV [42] which are one of LHC energies. Comparing with our recent
works [40, 41], we use different functions for $p_T$ and $y$
spectra in the present work, which reflects the flexibility of the
model and method used by us.

The rest part of this paper is structured as followings. The model
and method are concisely described in section 2. Results and
discussion are given in section 3. In section 4, we summarize our
main observations and conclusions.
\\

{\section{The model and method}}

The model used in the present work is a hybrid model, in which the
$p_T$ spectrum is described by a two-component distribution which
contains an inverse power-law suggested by the QCD calculus
[18--20] and an Erlang distribution resulted from a multisource
thermal model [17], and the $y$ spectrum is described by a
Gaussian distribution resulted from the Landau hydrodynamic model
[21--24]. The $\eta$ spectrum is also described due to the
Gaussian $y$ distribution and the two-component $p_T$
distribution, where the conversion between $y$ and $\eta$ is
accurately considered.

According to the QCD calculus [18--20], the $p_T$ spectrum in
high-$p_T$ region is described by the inverse power-law
\begin{equation}
f_1(p_{T})=Ap_T \bigg(1+\frac{p_T}{p_0} \bigg)^{-n},
\end{equation}
where $p_0$ and $n$ are free parameters, and $A$ is the
normalization constant which results in $\int_0^{\infty}
f_1(p_{T})dp_T=1$ and is related to the free parameters. According
to the multisource thermal model [17], the Erlang distribution
which describes the $p_T$ spectrum for a given sample is given by
\begin{equation}
f_2(p_{T})=\frac{p_T^{m-1}}{(m-1)!\langle p_{Ti} \rangle^m} \exp
\bigg(- \frac{p_{T}}{\langle p_{Ti} \rangle} \bigg),
\end{equation}
where $\langle p_{Ti} \rangle$ and $m$ are free parameters. We can
use $f_1(p_T)$ and $f_2(p_T)$ to describe the hard scattering
process and soft excitation process respectively. Let $k$ denote
the contribution ratio (relative contribution) of the hard
process, the final $p_T$ spectrum is described by the
two-component distribution
\begin{equation}
f_{p_T}(p_{T})=kf_1(p_T)+(1-k)f_2(p_T).
\end{equation}

According to the Landau hydrodynamic model [21--24], the
interacting system can be described by the hydrodynamics, which
results in the $y$ spectrum to be a Gaussian function [23, 24]
\begin{equation}
f_y(y)=\frac{1}{\sqrt{2\pi} \sigma_y} \exp \bigg[-
\frac{(y-y_C)^2}{2\sigma_y^2} \bigg],
\end{equation}
where $\sigma_y$ denotes the dispersion or width of rapidity
distribution and $y_C$ denotes the peak position or mid-rapidity.
In symmetric collisions such as in $pp$ and Pb-Pb collisions at
the LHC discussed in the present work, we have $y_C=0$ in the
laboratory or center-of-mass reference frame. The experimental
$\eta$ spectrum is also described by the Gaussian $y$ distribution
and the two-component $p_T$ distribution, in the case of the
conversion between $y$ and $\eta$ is accurately considered by a
Monte Carlo method. In some cases, the $y$ spectrum is described
by two Gaussian distributions, one is for the backward $y$ region
and the other one is for the forward $y$ region. In the case of
considering the three Gaussian distributions, the third one is for
the central $y$ region.

In the Monte Carlo method, let $R_{1,2,3,4}$ and $r_i$ ($i=1$, 2,
..., $m$) denote random numbers distributed evenly in [0,1]. We
have
\begin{equation}
\int_0^{p_T}f_1(p_T)dp_T<R_1<\int_0^{p_T+dp_T}f_1(p_T)dp_T
\end{equation}
due to Eq. (1), where $p_T$ in the upper limit of integral changes
from 0 to the maximum. Or
\begin{equation}
p_T=-\langle p_{Ti} \rangle \sum_{i=1}^m \ln r_i = -\langle p_{Ti}
\rangle \ln \prod_{i=1}^m r_i
\end{equation}
due to Eq. (2).
\begin{equation}
y=\sigma_y \sqrt{-2\ln R_2} \cos(2\pi R_3) +y_C
\end{equation}
due to Eq. (4). The azimuth $\varphi$ can be given by
\begin{equation}
\varphi=2\pi R_4
\end{equation}
due to $\varphi$ distributing evenly in $[0,2\pi]$ for an
isotropic source in the transverse plane.

In the considered reference frame such as the laboratory or
center-of-mass reference frame, the energy $E$ is given by
\begin{equation}
E=\sqrt{p_T^2+m_0^2}\cosh y,
\end{equation}
where $m_0$ denotes the rest mass of the considered particle. In
the case of considering unidentified charged particles, we take
$m_0=0.174$ GeV/$c^2$ which is estimated from an average weighted
the masses and yields of different types of charged particles [7].
The $x$-, $y$-, and $z$-components of momentum and velocity are
given by
\begin{equation}
p_x=p_T \cos \varphi, \quad p_y=p_T \sin \varphi, \quad
p_z=\sqrt{p_T^2+m_0^2}\sinh y,
\end{equation}
and
\begin{equation}
\beta_x=\frac{p_x}{E}, \quad \beta_y=\frac{p_y}{E}, \quad
\beta_z=\frac{p_z}{E}=\tanh y,
\end{equation}
respectively. The polar angle $\theta$ and the pseudorapidity
$\eta$ can be given by
\begin{equation}
\theta=\arctan \bigg( \frac{p_T}{p_z} \bigg)
\end{equation}
and
\begin{equation}
\eta \equiv -\ln\tan \bigg( \frac{\theta}{2} \bigg)
\end{equation}
respectively.

In the above discussions, a series of values of $\eta$ can be
obtained due to the Gaussian $y$ distribution (Eq. (4)) and the
two-component $p_T$ distribution (Eq. (3)), where the conversion
between $y$ and $\eta$ is accurately considered by the Monte Carlo
calculation. Then, the final $\eta$ distribution is obtained by
the statistics. At the same time, based on the Monte Carlo
calculation, a series of values of velocity and momentum
components can be obtained. Then, we can present and compare the
event patterns (particle scatter plots) in the three-dimensional
velocity and momentum spaces at the stage of kinetic freeze-out of
the interacting system for different particles produced in
different collisions at different energies, where different
particles may produce at different stages of collisions and carry
different information of interactions.
\\

{\section{Results and discussion}}

Figure 1 presents the transverse momentum spectra, $d^2\sigma/(p_T
d\eta dp_T)$, of charged particles produced in $pp$ collisions at
$\sqrt{s}=2.76$ TeV and in Pb-Pb collisions at
$\sqrt{s_{NN}}=2.76$ TeV in the pseudorapidity interval
$|\eta|<2$, where $\sigma$ denotes the cross-section, and the
integral luminosity $L_{\rm int}^{\rm pp}=4.2$ pb$^{-1}$ for $pp$
collisions and $L_{\rm int}^{\rm PbPb}=0.15$ nb$^{-1}$ for Pb-Pb
collisions. The symbols represent the experimental data of the
ATLAS Collaboration [42], where the data for Pb-Pb collisons are
divided by $\langle T_{AA} \rangle$ which is estimated as the
number of nucleon-nucleon collisions over their cross section [42,
43], and multiplied by different amounts marked in the panel. The
curves are our results calculated by using the two-component $p_T$
distribution (Eq. (3)). In the calculation, the method of least
squares is used to determine the values of parameters when we do
the fit to experimental data. The values of free parameters
($p_0$, $n$, $k$, $m$, and $\langle p_{Ti} \rangle$),
normalization constants ($N_{p_T}$), and $\chi^2$ per degree of
freedom ($\chi^2$/dof) are listed in Table 1, where the
normalization constant $N_{p_T}$ is used to give comparison
between the normalized curve with experimental data, and the
values of $m$ in the Erlang distribution are invariably taken to
be 2 which are not listed in the column. One can see that the
results calculated by using the hybrid model are in agreement with
the experimental $p_T$ data of charged particles produced in $pp$
and Pb-Pb collisions at 2.76 TeV measured by the ATLAS
Collaboration. The values of $p_0$, $n$, and $k$ for the inverse
power-law increase with the decrease of centrality (or with the
increase of centrality percentage), and the values of $\langle
p_{Ti} \rangle$ for the Erlang distribution does not show an
obvious tendency with the decrease of centrality. The
contributions of inverse power-law are not always main. We shall
discuss further the characteristics of parameters in the latter
part of this section.

Figures 2 and 3 are the same as those for Figure 1, but they show
the results in different $|\eta|$ intervals in $pp$ and 0--5\%
Pb-Pb collisions, respectively. The values of free parameters
($p_0$, $n$, $k$, $m$, and $\langle p_{Ti} \rangle$),
normalization constants ($N_{p_T}$), and $\chi^2$ per degree of
freedom ($\chi^2$/dof) are listed in Table 1, where the values of
$m$ in the Erlang distribution are invariably taken to be 2 which
are not listed in the column. One can see that the results
calculated by using the hybrid model are in agreement with the
experimental $p_T$ data of charged particles with different
$|\eta|$ intervals in $pp$ and 0--5\% Pb-Pb collisions at 2.76 TeV
measured by the ATLAS Collaboration. The values of $p_0$ increases
slightly and $\langle p_{Ti} \rangle$ decreases slightly with the
increase of $|\eta|$ in $pp$ collisions, and they do not show an
obvious tendency in 0--5\% Pb-Pb collisions. With the increase of
$|\eta|$, $n$ increases and $k$ does not change in both $pp$ and
0--5\% Pb-Pb collisions. Once again, the contributions of inverse
power-law are not always main.

\begin{figure}
\hskip-1.0cm \begin{center}
\includegraphics[width=12.0cm]{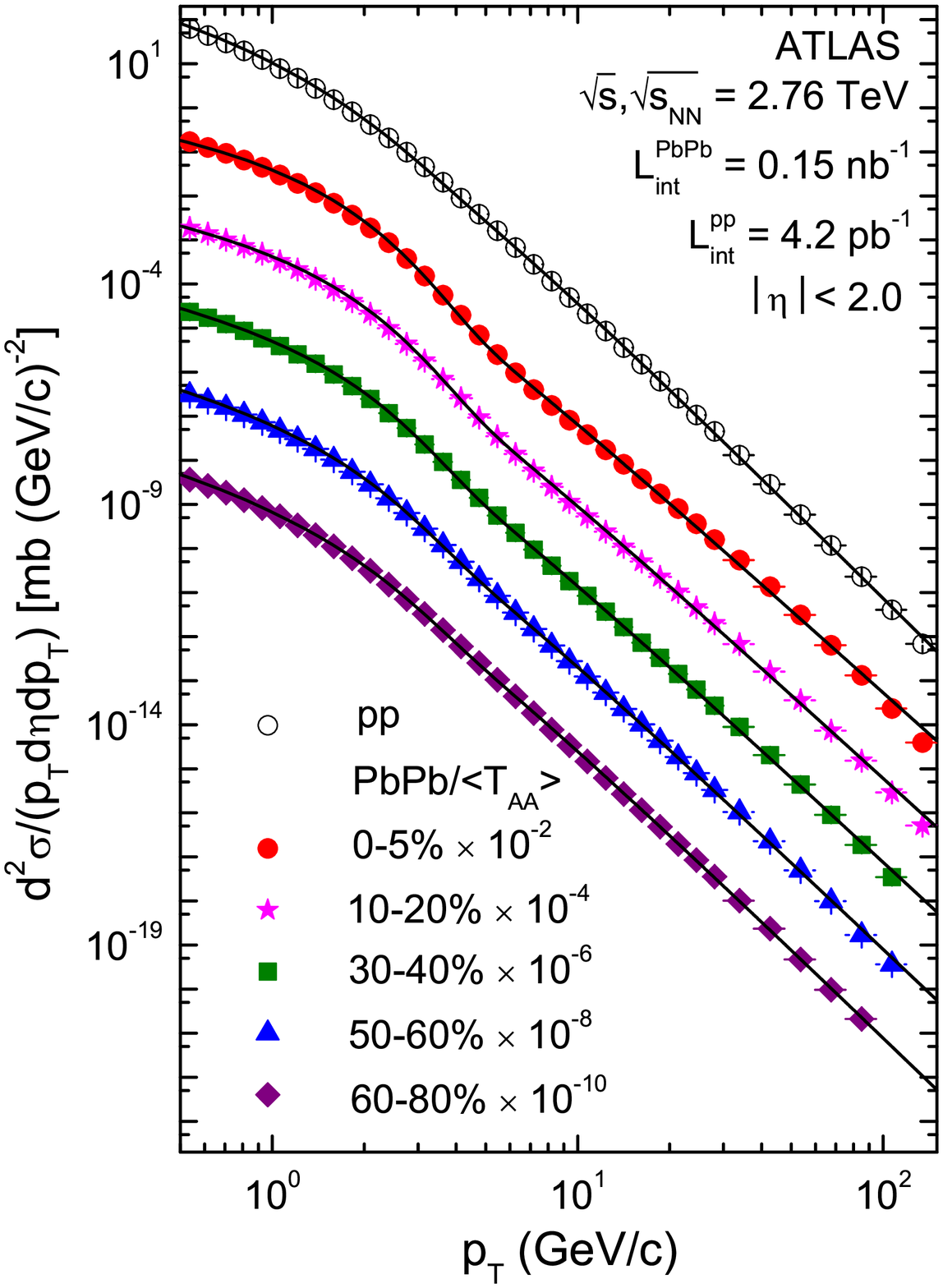}
\end{center}
\vskip.0cm Fig. 1. Transverse momentum spectra of charged
particles produced in $pp$ and Pb-Pb collisions at 2.76 TeV. The
symbols represent the data measured by the ATLAS Collaboration
[42] and the curves are our results calculated by using the
two-component distribution. For different centrality intervals,
the spectra are multiplied by different amounts marked in the
panels.
\end{figure}

\begin{figure}
\hskip-1.0cm \begin{center}
\includegraphics[width=12.0cm]{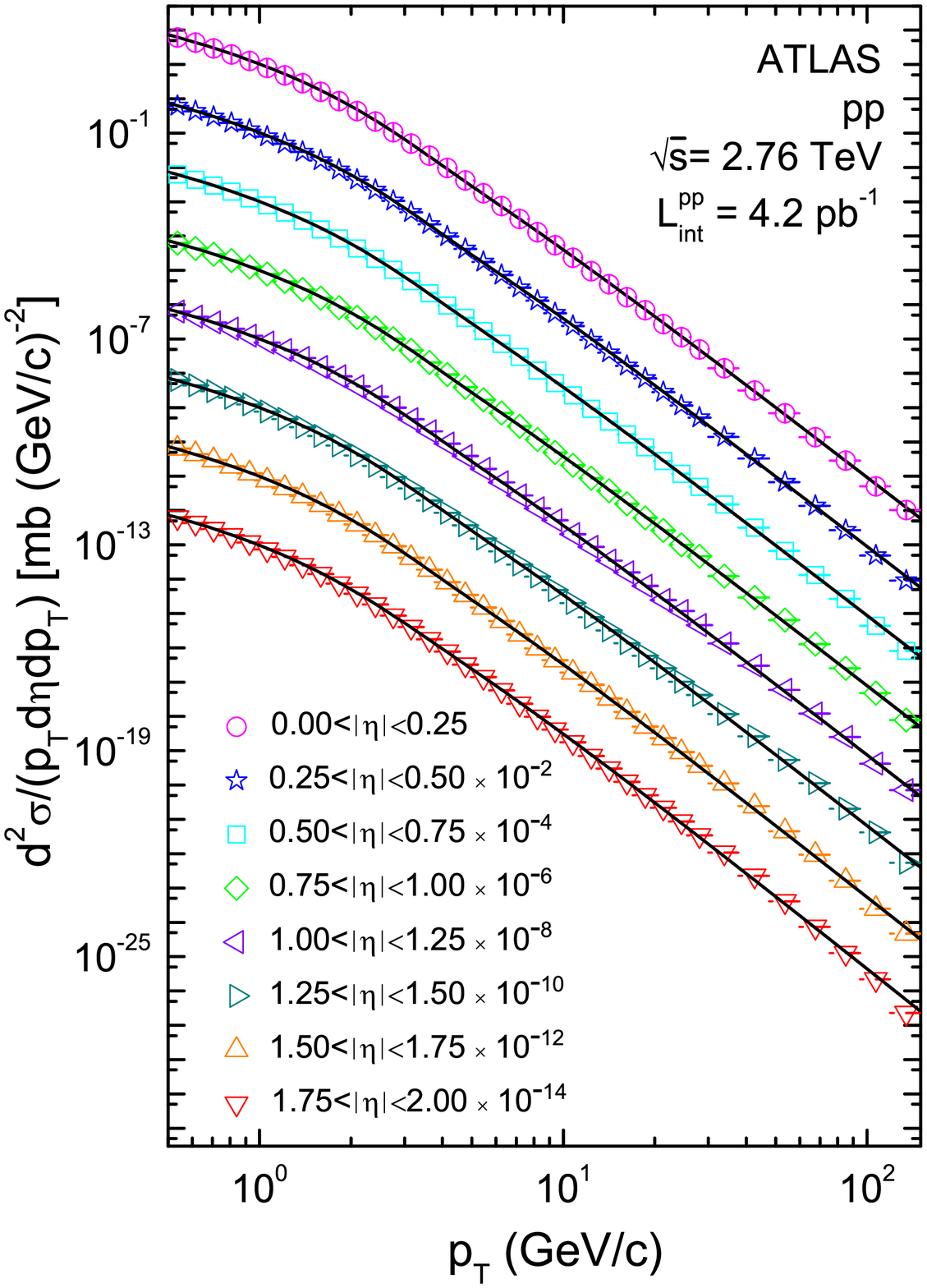}
\end{center}
\vskip.0cm Fig. 2. Same as Figure 1, but showing the results in
different pseudorapidity ranges in $pp$ collisions.
\end{figure}

\begin{figure}
\hskip-1.0cm \begin{center}
\includegraphics[width=12.0cm]{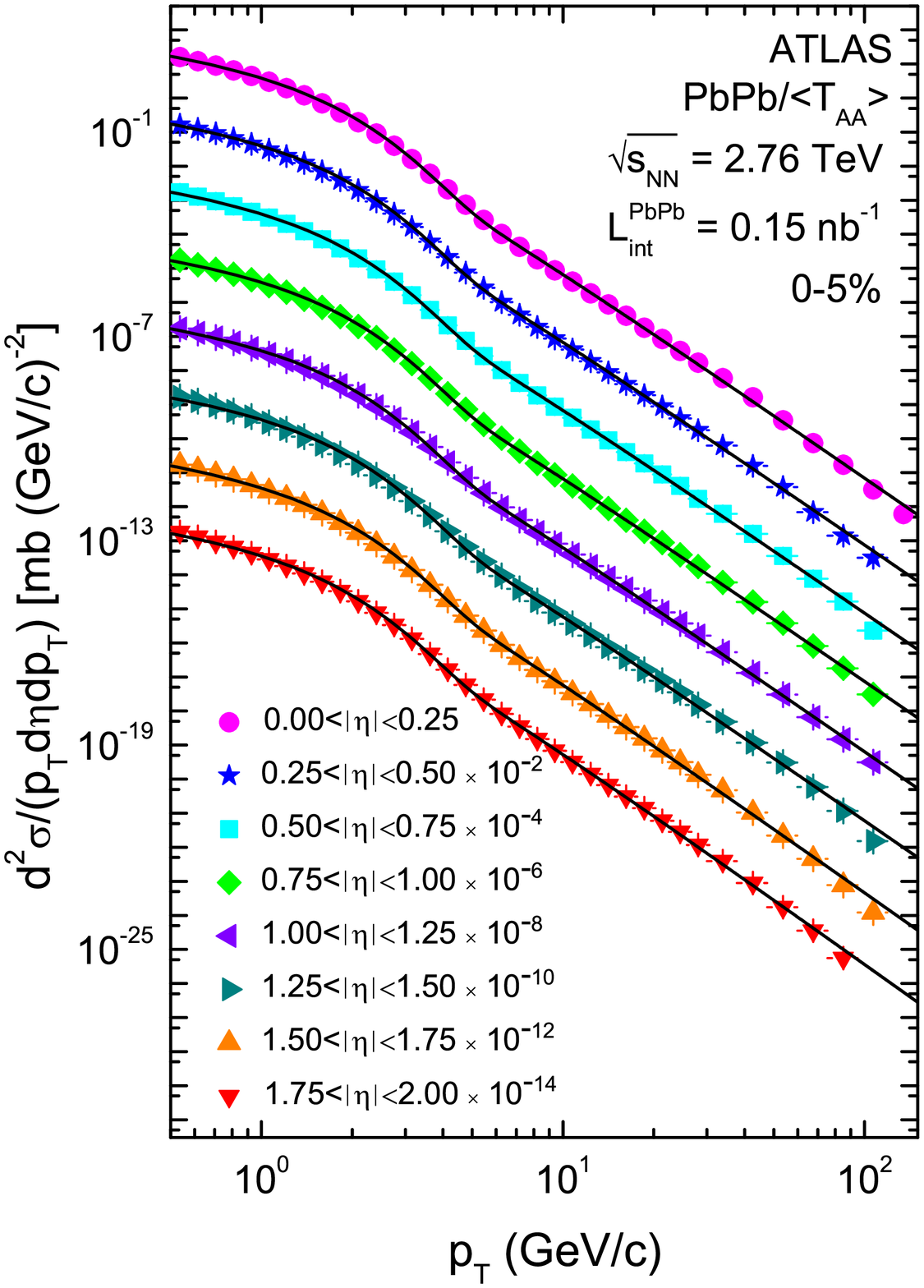}
\end{center}
\vskip.0cm Fig. 3. Same as Figure 1, but showing the results in
different pseudorapidity ranges in 0--5\% Pb-Pb collisions.
\end{figure}

\begin{figure}
\hskip-1.0cm \begin{center}
\includegraphics[width=16.0cm]{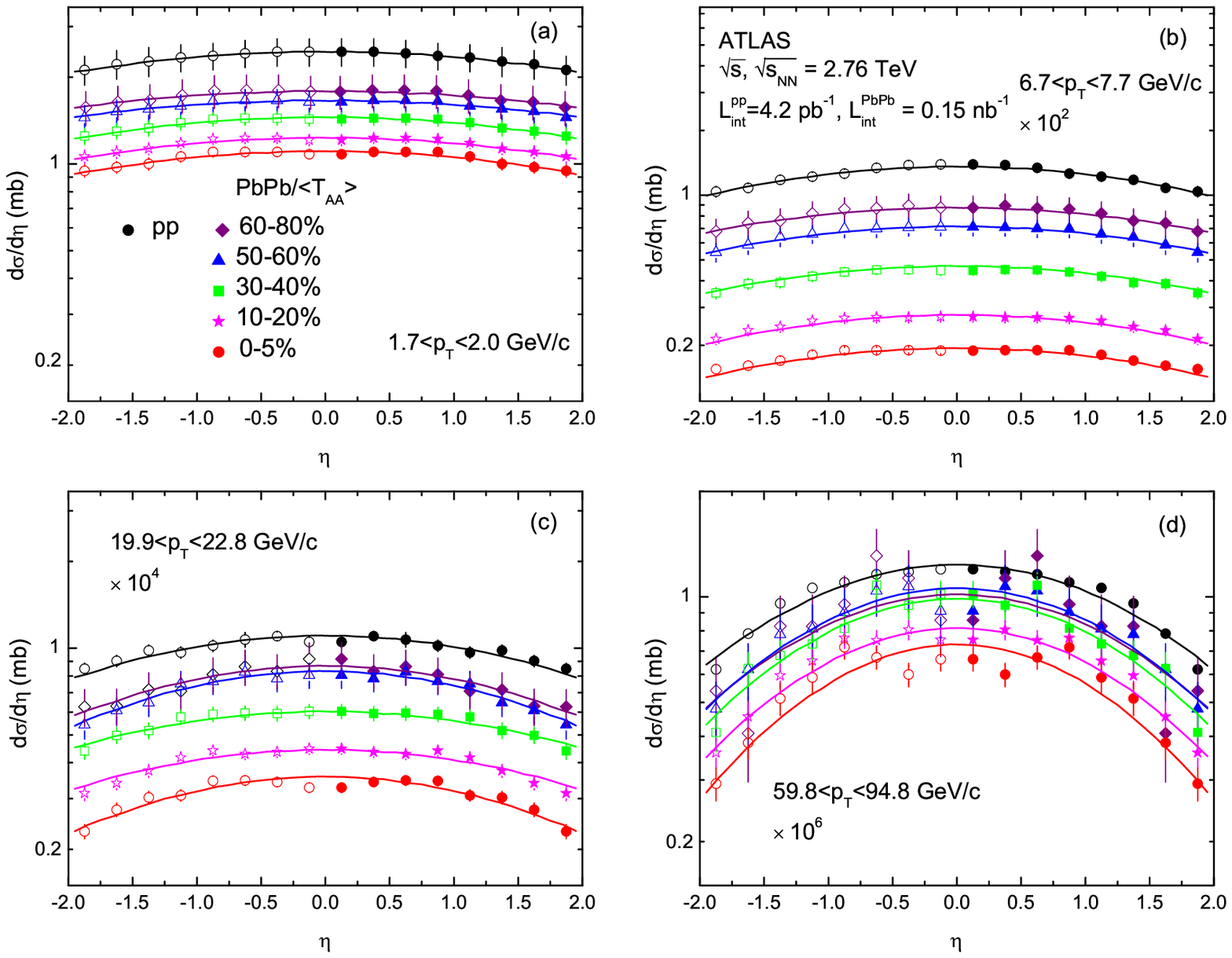}
\end{center}
\vskip.0cm Fig. 4. Pseudorapidity spectra of charged particles
produced in $pp$ and Pb-Pb collisions at 2.76 TeV for four $p_T$
intervals: (a) $1.7<p_T<2.0$ GeV/$c$, (b) $6.7<p_T<7.7$ GeV/$c$,
(c) $19.9<p_T<22.8$ GeV/$c$, and (d) $59.8<p_T<94.8$ GeV/$c$. The
symbols represent the data measured by the ATLAS Collaboration
[42] and the curves are our results calculated by using the
Gaussian $y$ distribution (and the two-component $p_T$
distribution), where the conversion between $y$ and $\eta$ is
considered. For different transverse momentum intervals, the
spectra are multiplied by different amounts marked in the panels.
\end{figure}

Based on the two-component $p_T$ distribution in which the
parameter values are obtained from Figure 1, and the Gaussian $y$
distribution in which the parameter values need to be determined,
we can perform the Monte Carlo calculation and obtain a series of
values of $\eta$. Thus, the $\eta$ distribution is obtained by the
statistics. Figure 4 shows the $\eta$ spectra of charged particles
produced in $pp$ and Pb-Pb collisions at 2.76 TeV for four $p_T$
intervals: (a) $1.7<p_T<2.0$ GeV/$c$, (b) $6.7<p_T<7.7$ GeV/$c$,
(c) $19.9<p_T<22.8$ GeV/$c$, and (d) $59.8<p_T<94.8$ GeV/$c$. The
symbols represent the experimental data measured by the ATLAS
Collaboration [42] and the curves are our results calculated by
the Gaussian $y$ distribution and the two-component $p_T$
distribution, where the conversion between $y$ and $\eta$ is
considered. For different $p_T$ intervals, the spectra are
multiplied by different amounts marked in the panels. The values
of free parameters ($\sigma_y$ and $y_C$), normalization constants
($N_{y}$), and $\chi^2$/dof are listed in Table 2, where the
normalization constant $N_{y}$ is used to give comparison between
the normalized curve with experimental data, and the values of
$y_C$ are not listed in the column due to $y_C=0$ at all time. One
can see that the results calculated by using the hybrid model are
approximately in agreement with the experimental $\eta$ data of
charged particles produced in $pp$ and Pb-Pb collisions at 2.76
TeV measured by the ATLAS Collaboration. The width of rapidity
distribution decreases with the increases of $p_T$ and centrality.
Although the Gaussian $y$ distribution in some cases has a space
to be extended to the two-Gaussian $y$ distribution [41], only the
Gaussian $y$ distribution is approximately used in the present
work due to a representation of the methodology.

Further, based on the parameter values obtained from Figures 1 and
4, we can perform the Monte Carlo calculation and obtain a series
of values of kinematical quantities. As a diagrammatic sketch,
Figure 5 presents the event patterns (particle scatter plots) in
the three-dimensional velocity ($\beta_x-\beta_y-\beta_z$) space
at the kinetic freeze-out of the interacting system formed in $pp$
collisions for four $p_T$ intervals: (a) $1.7<p_T<2.0$ GeV/$c$,
(b) $6.7<p_T<7.7$ GeV/$c$, (c) $19.9<p_T<22.8$ GeV/$c$, and (d)
$59.8<p_T<94.8$ GeV/$c$. The blue and red globules represent the
contributions of inverse power-law and Erlang distribution
respectively, where the red globules in the second $p_T$ intervals
are highlighted for clarity. The number of particles for each
panel is 1000. The values of root-mean-squares
($\sqrt{\overline{\beta_x^2}}$ for $\beta_x$,
$\sqrt{\overline{\beta_y^2}}$ for $\beta_y$, and
$\sqrt{\overline{\beta_z^2}}$ for $\beta_z$) and the maximum
$|\beta_x|$, $|\beta_y|$, and $|\beta_z|$ ($|\beta_x|_{\max}$,
$|\beta_y|_{\max}$, and $|\beta_z|_{\max}$) are listed in Table 3
which are obtained by higher statistics. The relative yields of
particle numbers appearing in different $p_T$ intervals are listed
in Table 4, where the relative yield in the highest $p_T$ interval
is taken to be 1. One can see that the contributions of inverse
power-law are main in the first two $p_T$ intervals, and sole in
the last two $p_T$ intervals. The contribution of Erlang
distribution can be neglected in the second $p_T$ interval. The
relations $\sqrt{\overline{\beta_x^2}} \approx
\sqrt{\overline{\beta_y^2}} \ll \sqrt{\overline{\beta_z^2}}$ and
$|\beta_x|_{\max} \approx |\beta_y|_{\max} \approx
|\beta_z|_{\max} \approx 1$ render that the root-mean-square
velocities form an ellipsoid surface with the major axis along the
beam direction, and the maximum velocities form a spherical
surface.

By using the same method as that for Figure 5, we can obtain the
similar results in Pb-Pb collisions with centrality intervals
60--80\%, 50--60\%, 30--40\%, 10--20\%, and 0--5\%, respectively.
The values of root-mean squares of velocity components and the
maximum velocity components are listed in Table 3, and the
relative yields of particle numbers appearing in different $p_T$
intervals are listed in Table 4. As an example, to reduce the size
of the paper file, only the scatter plots in the three-dimensional
velocity space in 0--5\% Pb-Pb collisions are given in Figure 6.
Some conclusions obtained from Figures 1 and 5 can be obtained
from Figure 6 and Tables 3 and 4. In addition, we see intuitively
the density change of particle numbers in the three-dimensional
velocity space in different $p_T$ intervals at the kinetic
freeze-out of the interacting system formed in Pb-Pb collisions
with different centrality intervals.

\newpage {\scriptsize {Table 1. Values of free parameters ($p_{0}$,
$n$, $k$, $m$, and $\langle p_{Ti} \rangle$), normalization
constant ($N_{p_T}$), and $\chi^2$/dof corresponding to the curves
in Figures 1--3, where the values of $m$ in the Erlang
distribution are invariably taken to be 2 which are not listed in
the column.
{%
\begin{center}
\begin{tabular}{cccccccc}
\hline\hline Figure & Type & $p_{0}$ (GeV/$c$) & $n$ & $k$ & $\langle p_{Ti} \rangle$ (GeV/$c$) & $N_{p_T}$ & $\chi^2$/dof \\
\hline
Figure 1 & $pp$                & $0.82\pm0.04$ & $6.90\pm0.10$ & $0.90\pm0.05$ & $0.37\pm0.02$ & $345.0\pm17.3$   & $6.634$\\
         & PbPb, C = 0--5\%    & $0.66\pm0.03$ & $6.24\pm0.10$ & $0.41\pm0.05$ & $0.38\pm0.02$ & $72.3\pm3.6$     & $12.476$\\
         & PbPb, C = 10--20\%  & $0.70\pm0.04$ & $6.34\pm0.10$ & $0.50\pm0.05$ & $0.39\pm0.02$ & $83.8\pm4.2$     & $10.719$\\
         & PbPb, C = 30--40\%  & $0.72\pm0.04$ & $6.44\pm0.10$ & $0.63\pm0.05$ & $0.39\pm0.02$ & $115.5\pm5.8$    & $4.926$\\
         & PbPb, C = 50--60\%  & $0.75\pm0.04$ & $6.59\pm0.10$ & $0.76\pm0.05$ & $0.39\pm0.02$ & $164.1\pm8.2$    & $1.686$\\
         & PbPb, C = 60--80\%  & $0.78\pm0.04$ & $6.70\pm0.10$ & $0.85\pm0.05$ & $0.40\pm0.02$ & $202.2\pm10.1$   & $0.681$\\
\hline
Figure 2 & $pp$, $0.00<|\eta|<0.25$ & $0.85\pm0.04$ & $6.85\pm0.10$ & $0.87\pm0.05$ & $0.39\pm0.02$ & $304.0\pm15.2$ & $5.458$\\
         & $pp$, $0.25<|\eta|<0.50$ & $0.85\pm0.04$ & $6.86\pm0.10$ & $0.87\pm0.05$ & $0.39\pm0.02$ & $306.8\pm15.3$ & $5.575$\\
         & $pp$, $0.50<|\eta|<0.75$ & $0.85\pm0.04$ & $6.87\pm0.10$ & $0.88\pm0.05$ & $0.39\pm0.02$ & $304.3\pm15.2$ & $6.399$\\
         & $pp$, $0.75<|\eta|<1.00$ & $0.86\pm0.04$ & $6.90\pm0.10$ & $0.88\pm0.05$ & $0.39\pm0.02$ & $299.6\pm15.0$ & $6.355$\\
         & $pp$, $1.00<|\eta|<1.25$ & $0.86\pm0.04$ & $6.91\pm0.10$ & $0.87\pm0.05$ & $0.38\pm0.02$ & $301.1\pm15.1$ & $6.218$\\
         & $pp$, $1.25<|\eta|<1.50$ & $0.87\pm0.04$ & $6.95\pm0.10$ & $0.87\pm0.05$ & $0.38\pm0.02$ & $295.9\pm14.8$ & $5.127$\\
         & $pp$, $1.50<|\eta|<1.75$ & $0.88\pm0.05$ & $7.01\pm0.10$ & $0.87\pm0.05$ & $0.37\pm0.02$ & $295.2\pm14.8$ & $5.675$\\
         & $pp$, $1.75<|\eta|<2.00$ & $0.90\pm0.05$ & $7.08\pm0.10$ & $0.88\pm0.05$ & $0.35\pm0.02$ & $297.1\pm14.9$ & $6.574$\\
\hline
Figure 3 & PbPb, $0.00<|\eta|<0.25$ & $0.67\pm0.04$ & $6.13\pm0.10$ & $0.36\pm0.05$ & $0.39\pm0.02$ & $67.3\pm3.4$ & $12.044$\\
         & PbPb, $0.25<|\eta|<0.50$ & $0.65\pm0.04$ & $6.10\pm0.10$ & $0.35\pm0.05$ & $0.39\pm0.02$ & $67.5\pm3.4$ & $12.713$\\
         & PbPb, $0.50<|\eta|<0.75$ & $0.65\pm0.03$ & $6.10\pm0.10$ & $0.35\pm0.05$ & $0.39\pm0.02$ & $67.6\pm3.4$ & $12.566$\\
         & PbPb, $0.75<|\eta|<1.00$ & $0.65\pm0.03$ & $6.10\pm0.10$ & $0.35\pm0.05$ & $0.39\pm0.02$ & $67.4\pm3.4$ & $10.751$\\
         & PbPb, $1.00<|\eta|<1.25$ & $0.66\pm0.03$ & $6.14\pm0.10$ & $0.35\pm0.05$ & $0.39\pm0.02$ & $66.2\pm3.3$ & $7.959$\\
         & PbPb, $1.25<|\eta|<1.50$ & $0.66\pm0.03$ & $6.14\pm0.10$ & $0.35\pm0.05$ & $0.39\pm0.02$ & $62.3\pm3.1$ & $8.215$\\
         & PbPb, $1.50<|\eta|<1.75$ & $0.68\pm0.03$ & $6.26\pm0.10$ & $0.39\pm0.05$ & $0.39\pm0.02$ & $62.7\pm3.1$ & $9.323$\\
         & PbPb, $1.75<|\eta|<2.00$ & $0.70\pm0.03$ & $6.33\pm0.10$ & $0.38\pm0.05$ & $0.38\pm0.02$ & $64.1\pm3.2$ & $7.913$\\
\hline
\end{tabular}%
\end{center}
}} }

\vskip1.0cm {\scriptsize {Table 2. Values of free parameter
($\sigma_y$ and $y_C$), normalization constant ($N_\eta$), and
$\chi^2$/dof corresponding to the curves in Figure 4, where the
values of $y_C$ are not listed in the column due to $y_C=0$ at all
time.
{%
\begin{center}
\begin{tabular}{cccccc}
\hline\hline  Figure & Type & $\sigma_y$ & $N_{\eta}$ & $\chi^2$/dof \\
\hline
                          & $pp$           & $3.40\pm0.17$ & $92.80\pm4.64$ & $0.003$ \\
                          & PbPb, 60--80\% & $3.80\pm0.19$ & $68.40\pm3.42$ & $0.037$ \\
Figure 4(a)               & PbPb, 50--60\% & $3.80\pm0.19$ & $63.50\pm3.18$ & $0.046$ \\
$1.7<p_{T}<2.0$ GeV/$c$   & PbPb, 30--40\% & $3.30\pm0.17$ & $54.80\pm2.74$ & $0.137$ \\
                          & PbPb, 10--20\% & $3.30\pm0.17$ & $46.50\pm2.33$ & $0.518$ \\
                          & PbPb,  0--5\%  & $3.20\pm0.16$ & $41.60\pm2.08$ & $0.661$ \\
\hline
                          & $pp$           & $2.50\pm0.13$ & $0.49\pm0.02$  & $0.687$ \\
                          & PbPb, 60--80\% & $2.70\pm0.14$ & $0.32\pm0.02$  & $0.086$ \\
Figure 4(b)               & PbPb, 50--60\% & $2.60\pm0.13$ & $0.26\pm0.01$  & $0.142$ \\
$6.7<p_{T}<7.7$ GeV/$c$   & PbPb, 30--40\% & $2.60\pm0.13$ & $0.17\pm0.01$  & $1.324$ \\
                          & PbPb, 10--20\% & $2.50\pm0.13$ & $0.10\pm0.01$  & $1.116$ \\
                          & PbPb,  0--5\%  & $2.50\pm0.13$ & $0.07\pm0.01$  & $1.856$ \\
\hline
                          & $pp$           & $2.40\pm0.12$ & $(3.95\pm0.20)\times10^{-3}$ & $1.643$ \\
                          & PbPb, 60--80\% & $2.20\pm0.11$ & $(3.04\pm0.15)\times10^{-3}$ & $0.461$ \\
Figure 4(c)               & PbPb, 50--60\% & $2.10\pm0.11$ & $(2.88\pm0.14)\times10^{-3}$ & $0.469$ \\
$19.9<p_{T}<22.8$ GeV/$c$ & PbPb, 30--40\% & $2.30\pm0.12$ & $(2.19\pm1.11)\times10^{-3}$ & $1.002$ \\
                          & PbPb, 10--20\% & $2.10\pm0.11$ & $(1.59\pm0.08)\times10^{-3}$ & $3.064$ \\
                          & PbPb,  0--5\%  & $2.01\pm0.10$ & $(1.24\pm0.06)\times10^{-3}$ & $4.665$ \\
\hline
                          & $pp$           & $1.70\pm0.09$ & $(4.00\pm0.20)\times10^{-5}$ & $3.310$ \\
                          & PbPb, 60--80\% & $1.60\pm0.08$ & $(3.21\pm0.16)\times10^{-5}$ & $3.865$ \\
Figure 4(d)               & PbPb, 50--60\% & $1.55\pm0.08$ & $(3.30\pm0.17)\times10^{-5}$ & $0.813$ \\
$59.8<p_{T}<94.8$ GeV/$c$ & PbPb, 30--40\% & $1.52\pm0.08$ & $(3.05\pm0.15)\times10^{-5}$ & $2.829$ \\
                          & PbPb, 10--20\% & $1.50\pm0.08$ & $(2.50\pm0.13)\times10^{-5}$ & $2.718$ \\
                          & PbPb,  0--5\%  & $1.40\pm0.07$ & $(2.17\pm0.11)\times10^{-5}$ & $6.708$ \\
\hline
\end{tabular}%
\end{center}
}} } \vskip1.0cm

Figure 7 presents the event patterns (particle scatter plots) in
the three-dimensional momentum ($p_x-p_y-p_z$) space at the
kinetic freeze-out of the interacting system formed in $pp$
collisions for four $p_T$ intervals: (a)(b) $1.7<p_T<2.0$ GeV/$c$,
(c)(d) $6.7<p_T<7.7$ GeV/$c$, (e)(f) $19.9<p_T<22.8$ GeV/$c$, and
(g)(h) $59.8<p_T<94.8$ GeV/$c$. The left and right panels display
the results in a wide (from $-1.5$ to 1.5 TeV$/c$) and narrow
(from $-50$ to 50 GeV$/c$) $p_z$ ranges respectively. The blue and
red globules represent the contributions of inverse power-law and
Erlang distribution respectively, where the red globules in the
second $p_T$ interval are highlighted for clarity. The number of
particles for each panel in the left is 1000. The values of
root-mean-squares ($\sqrt{\overline{p_x^2}}$ for $p_x$,
$\sqrt{\overline{p_y^2}}$ for $p_y$, and $\sqrt{\overline{p_z^2}}$
for $p_z$) and the maximum $|p_x|$, $|p_y|$, and $|p_z|$
($|p_x|_{\max}$, $|p_y|_{\max}$, and $|p_z|_{\max}$) are listed in
Table 5. The relative yields of particle numbers appearing in
different $p_T$ intervals are listed in Table 4, and the
percentages of particle numbers in the interval $-50<p_z<50$
GeV/$c$ over whole $p_z$ range for different $p_{T}$ intervals are
listed in Table 6, where ``PL" in Table 6 denotes the
``power-law". One can see that some conclusions obtained from
Figures 5 and 6 can be obtained from Figure 7. In the wide $p_z$
range, most particles constitute a circle-like region surrounded
by a few particles. In the narrow $p_z$ range, particles
constitute a cylinder surface surrounded by a few particles. We
have the relations $\sqrt{\overline{p_x^2}} \approx
\sqrt{\overline{p_y^2}} \ll \sqrt{\overline{p_z^2}}$ and
$|p_x|_{\max} \approx |p_y|_{\max} \ll |p_z|_{\max}$.

\begin{figure}
\hskip-1.0cm \begin{center}
\includegraphics[width=14.0cm]{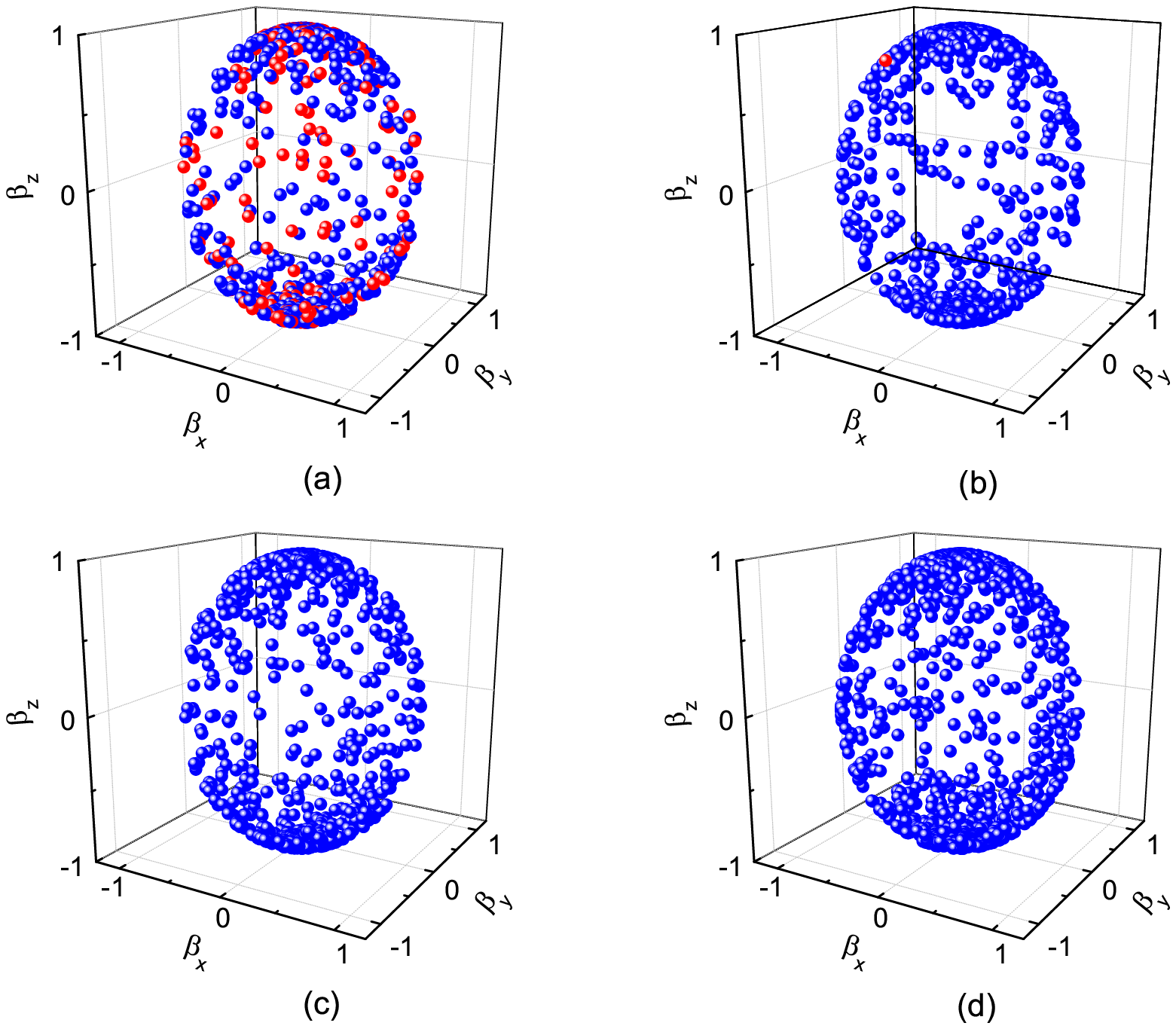}
\end{center}
\vskip.0cm Fig. 5. Event patterns (particle scatter plots) in
three-dimensional velocity ($\beta_x-\beta_y-\beta_z$) space at
kinetic freeze-out in $pp$ collisions for four transverse momentum
intervals: (a) $1.7<p_T<2.0$ GeV/$c$, (b) $6.7<p_T<7.7$ GeV/$c$,
(c) $19.9<p_T<22.8$ GeV/$c$, and (d) $59.8<p_T<94.8$ GeV/$c$. The
velocity components are in the units of $c$. The blue and red
globules represent the contributions of inverse power-law and
Erlang distribution respectively. The number of particles for each
panel is 1000.
\end{figure}

\begin{figure}
\hskip-1.0cm \begin{center}
\includegraphics[width=14.0cm]{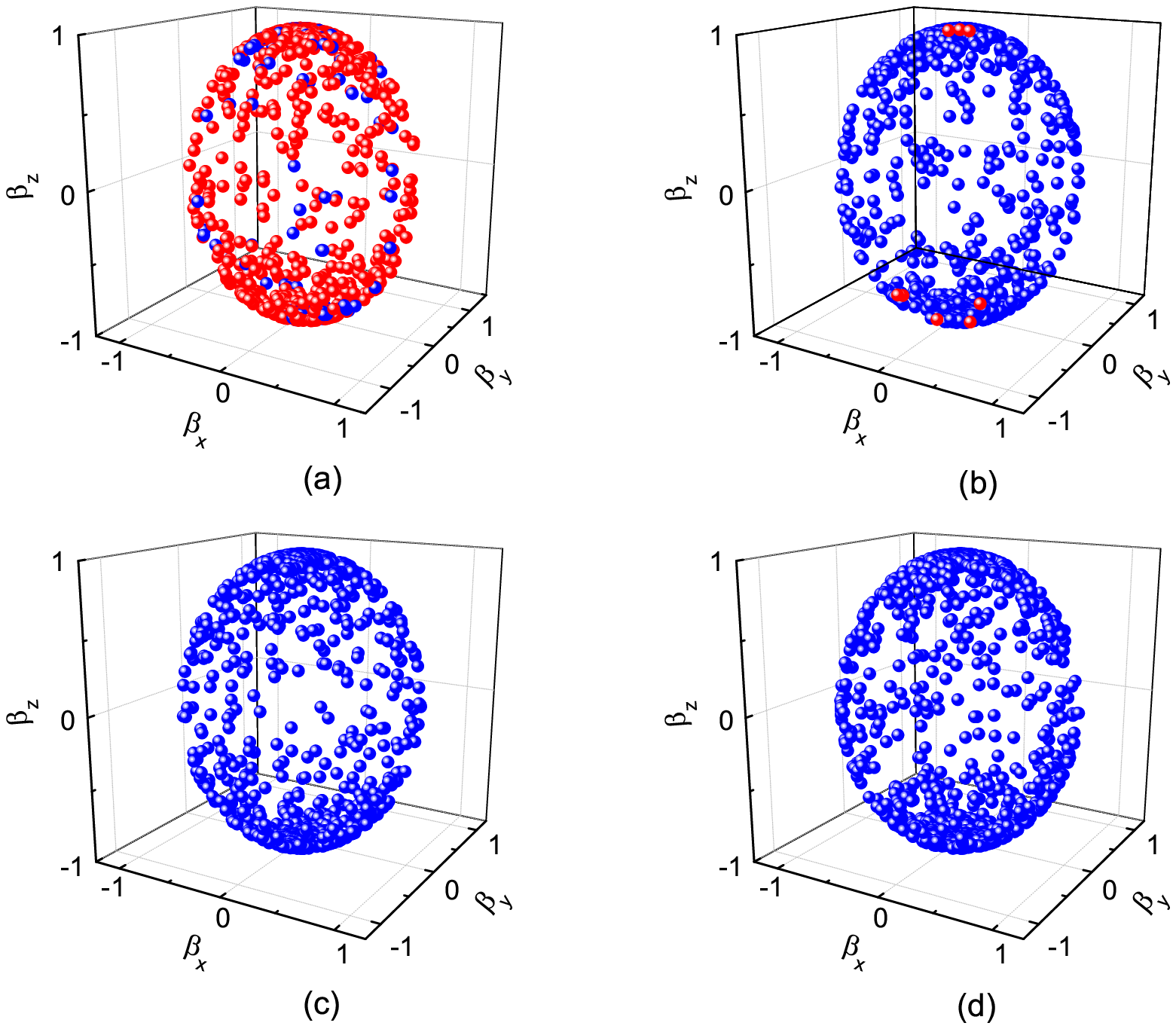}
\end{center}
\vskip.0cm Fig. 6. Same as Figure 5, but showing the results in
0--5\% Pb-Pb collisions.
\end{figure}

\renewcommand{\baselinestretch}{.5}
\begin{figure}
\hskip-1.0cm \begin{center}
\includegraphics[width=14.0cm]{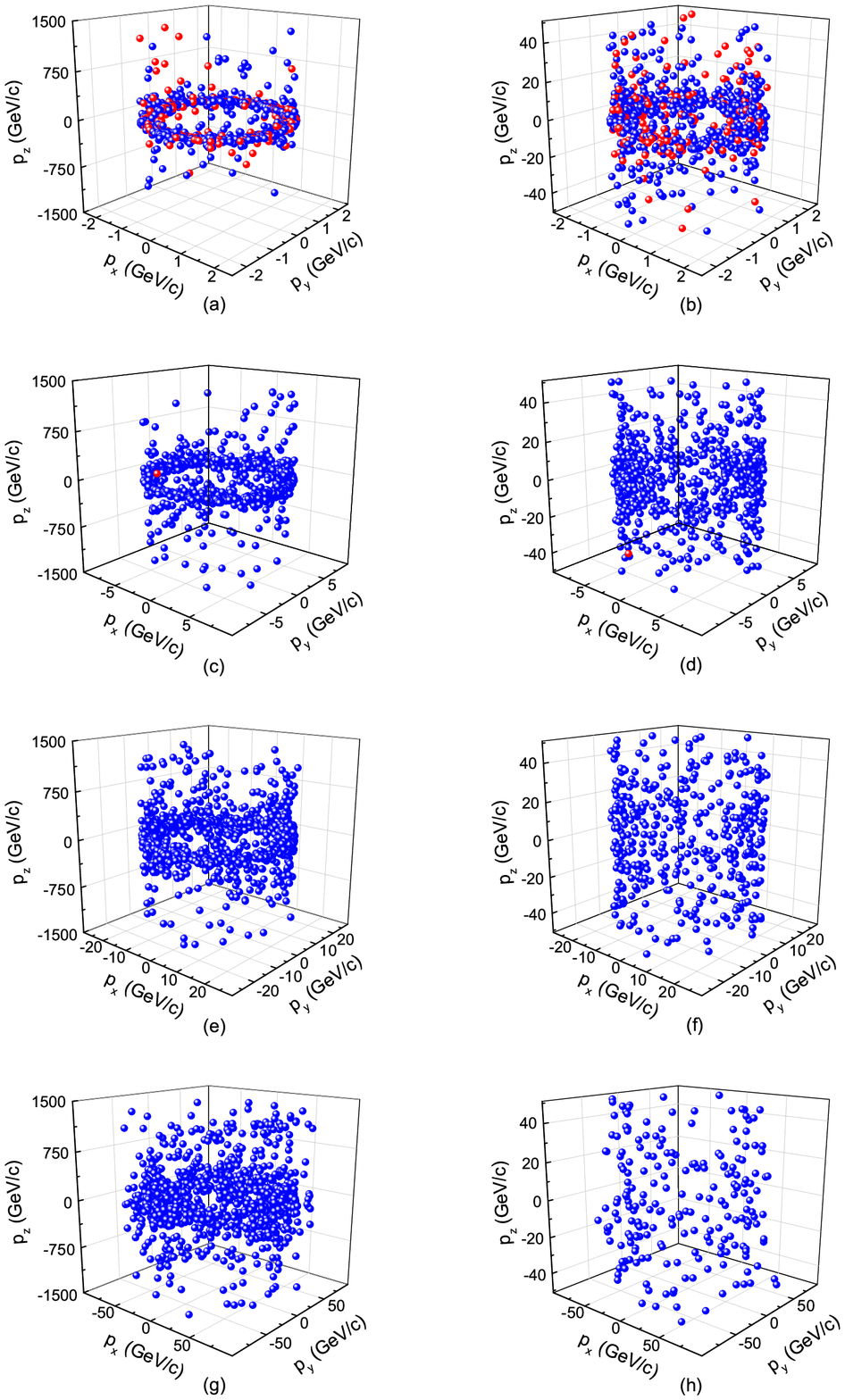}
\end{center}
\vskip.0cm {\scriptsize Fig. 7. Event patterns (particle scatter
plots) in three-dimensional momentum ($p_x-p_y-p_z$) space at
kinetic freeze-out in $pp$ collisions for four transverse momentum
intervals: (a)(b) $1.7<p_T<2.0$ GeV/$c$, (c)(d) $6.7<p_T<7.7$
GeV/$c$, (e)(f) $19.9<p_T<22.8$ GeV/$c$, and (g)(h)
$59.8<p_T<94.8$ GeV/$c$. The left and right panels display the
results in a wide and narrow $p_z$ ranges respectively. The blue
and red globules represent the contributions of inverse power-law
and Erlang distribution respectively. The number of particles for
each panel in the left is 1000.}
\end{figure}
\renewcommand{\baselinestretch}{.98}

\begin{figure}
\hskip-1.0cm \begin{center}
\includegraphics[width=14.0cm]{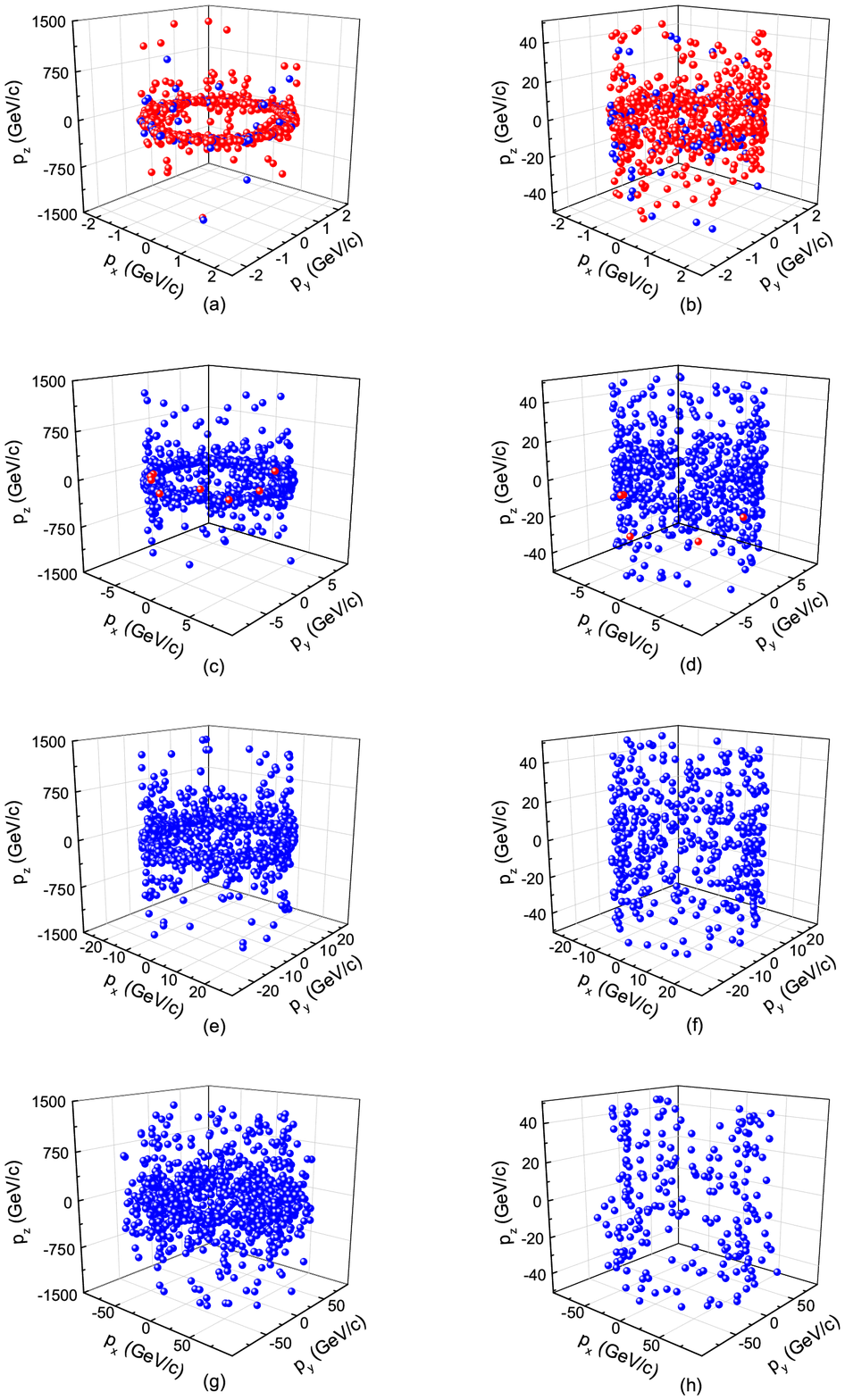}
\end{center}
\vskip.0cm Fig. 8. Same as Figure 7, but showing the results in
0--5\% Pb-Pb collisions.
\end{figure}

\vskip1.0cm {\scriptsize {Table 3. Values of the root-mean-squares
$\sqrt{\overline{\beta_x^2}}$ for $\beta_x$,
$\sqrt{\overline{\beta_y^2}}$ for $\beta_y$, and
$\sqrt{\overline{\beta_z^2}}$ for $\beta_z$, as well as the
maximum $|\beta_x|$, $|\beta_y|$, and $|\beta_z|$
($|\beta_x|_{\max}$, $|\beta_y|_{\max}$, and $|\beta_z|_{\max}$)
corresponding to the scatter plots in different types of
collisions, where the corresponding scatter plots in $pp$ and
0--5\% Pb-Pb collisions are presented in Figures 5 and 6
respectively. Both the root-mean-squares and the maximum velocity
components are in the units of $c$, and all the $p_T$ intervals
are in the units of GeV/$c$.
{%
\begin{center}
\begin{tabular}{ccccccc}
\hline\hline Type & $\sqrt{\overline{\beta_x^2}}$ & $\sqrt{\overline{\beta_y^2}}$ & $\sqrt{\overline{\beta_z^2}}$ & $|\beta_x|_{\max}$ & $|\beta_y|_{\max}$ & $|\beta_z|_{\max}$ \\
\hline
$pp$ & & & & & & \\
$1.7<p_{T}<2.0$   & $0.328\pm0.010$ & $0.343\pm0.010$ & $0.879\pm0.006$ & $0.993$ & $0.994$ & $1.000$ \\
$6.7<p_{T}<7.7$   & $0.359\pm0.010$ & $0.343\pm0.010$ & $0.868\pm0.006$ & $0.995$ & $0.997$ & $1.000$ \\
$19.9<p_{T}<22.8$ & $0.361\pm0.010$ & $0.354\pm0.010$ & $0.863\pm0.006$ & $1.000$ & $0.996$ & $1.000$ \\
$59.8<p_{T}<94.8$ & $0.413\pm0.009$ & $0.391\pm0.009$ & $0.823\pm0.007$ & $0.999$ & $0.999$ & $0.999$ \\
\hline
PbPb, 60--80\% & & & & & & \\
$1.7<p_{T}<2.0$   & $0.331\pm0.010$ & $0.336\pm0.010$ & $0.880\pm0.006$ & $0.992$ & $0.994$ & $1.000$ \\
$6.7<p_{T}<7.7$   & $0.358\pm0.010$ & $0.347\pm0.010$ & $0.867\pm0.006$ & $0.995$ & $0.997$ & $1.000$ \\
$19.9<p_{T}<22.8$ & $0.368\pm0.010$ & $0.332\pm0.009$ & $0.869\pm0.006$ & $1.000$ & $0.995$ & $1.000$ \\
$59.8<p_{T}<94.8$ & $0.405\pm0.009$ & $0.390\pm0.009$ & $0.827\pm0.006$ & $0.998$ & $0.995$ & $0.999$ \\
\hline
PbPb, 50--60\% & & & & & & \\
$1.7<p_{T}<2.0$   & $0.324\pm0.010$ & $0.331\pm0.010$ & $0.885\pm0.006$ & $0.992$ & $0.994$ & $1.000$ \\
$6.7<p_{T}<7.7$   & $0.354\pm0.010$ & $0.351\pm0.010$ & $0.866\pm0.006$ & $0.995$ & $0.997$ & $1.000$ \\
$19.9<p_{T}<22.8$ & $0.364\pm0.010$ & $0.334\pm0.009$ & $0.869\pm0.006$ & $1.000$ & $0.988$ & $1.000$ \\
$59.8<p_{T}<94.8$ & $0.399\pm0.009$ & $0.386\pm0.009$ & $0.832\pm0.006$ & $0.998$ & $1.000$ & $0.999$ \\
\hline
PbPb, 30--40\% & & & & & & \\
$1.7<p_{T}<2.0$   & $0.325\pm0.009$ & $0.350\pm0.010$ & $0.877\pm0.006$ & $0.993$ & $0.994$ & $1.000$ \\
$6.7<p_{T}<7.7$   & $0.362\pm0.010$ & $0.344\pm0.010$ & $0.866\pm0.006$ & $0.995$ & $0.997$ & $1.000$ \\
$19.9<p_{T}<22.8$ & $0.374\pm0.010$ & $0.358\pm0.010$ & $0.856\pm0.006$ & $1.000$ & $0.996$ & $1.000$ \\
$59.8<p_{T}<94.8$ & $0.414\pm0.009$ & $0.389\pm0.009$ & $0.823\pm0.006$ & $0.995$ & $0.999$ & $0.999$ \\
\hline
PbPb, 10--20\% & & & & & & \\
$1.7<p_{T}<2.0$   & $0.325\pm0.009$ & $0.350\pm0.010$ & $0.877\pm0.006$ & $0.993$ & $0.994$ & $1.000$ \\
$6.7<p_{T}<7.7$   & $0.362\pm0.010$ & $0.344\pm0.010$ & $0.866\pm0.006$ & $0.995$ & $0.997$ & $1.000$ \\
$19.9<p_{T}<22.8$ & $0.374\pm0.010$ & $0.358\pm0.010$ & $0.856\pm0.006$ & $1.000$ & $0.996$ & $1.000$ \\
$59.8<p_{T}<94.8$ & $0.413\pm0.009$ & $0.389\pm0.009$ & $0.823\pm0.006$ & $0.995$ & $0.999$ & $0.999$ \\
\hline
PbPb, 0--5\% & & & & & & \\
$1.7<p_{T}<2.0$   & $0.343\pm0.010$ & $0.353\pm0.010$ & $0.869\pm0.006$ & $0.993$ & $0.992$ & $1.000$ \\
$6.7<p_{T}<7.7$   & $0.369\pm0.010$ & $0.348\pm0.010$ & $0.862\pm0.006$ & $0.995$ & $0.997$ & $1.000$ \\
$19.9<p_{T}<22.8$ & $0.378\pm0.010$ & $0.356\pm0.009$ & $0.855\pm0.006$ & $1.000$ & $0.995$ & $1.000$ \\
$59.8<p_{T}<94.8$ & $0.419\pm0.009$ & $0.388\pm0.009$ & $0.821\pm0.007$ & $0.995$ & $0.999$ & $0.999$ \\
\hline
\end{tabular}
\end{center}
}} }

\vskip2.0cm {\scriptsize {Table 4. Relative yields of particle
numbers appearing in different $p_T$ intervals in different types
of collisions, where the corresponding scatter plots in $pp$ and
0--5\% Pb-Pb collisions are presented in Figures 5 and 6, as well
as 7 and 8, respectively. The relative yields in the highest $p_T$
interval are taken to be 1. All the $p_T$ intervals are in the
units of GeV/$c$.
{%
\begin{center}
\begin{tabular}{ccccc}
\hline\hline
Type & $1.7<p_T<2.0$ & $6.7<p_T<7.7$ & $19.9<p_T<22.8$ & $59.8<p_T<94.8$ \\
\hline
$pp$           & $2.75\times10^{6}$ & $1.29\times10^{4}$ & 99.6 & $1$ \\
PbPb, 60--80\% & $2.04\times10^{6}$ & $8.61\times10^{3}$ & 79.6 & $1$ \\
PbPb, 50--60\% & $1.55\times10^{6}$ & $6.61\times10^{3}$ & 69.5 & $1$ \\
PbPb, 30--40\% & $1.91\times10^{6}$ & $5.08\times10^{3}$ & 59.4 & $1$ \\
PbPb, 10--20\% & $2.15\times10^{6}$ & $4.15\times10^{3}$ & 53.0 & $1$ \\
PbPb, 0--5\%   & $2.25\times10^{6}$ & $3.42\times10^{3}$ & 47.5 & $1$ \\
\hline
\end{tabular}
\end{center}
}} }

\vskip1.0cm {\scriptsize {Table 5. Values of the root-mean-squares
$\sqrt{\overline{p_x^2}}$ for $p_x$, $\sqrt{\overline{p_y^2}}$ for
$p_y$, and $\sqrt{\overline{p_z^2}}$ for $p_z$, as well as the
maximum $|p_x|$, $|p_y|$, and $|p_z|$ ($|p_x|_{\max}$,
$|p_y|_{\max}$, and $|p_z|_{\max}$) corresponding to the scatter
plots in different types of collisions, where the corresponding
scatter plots in $pp$ and 0--5\% Pb-Pb collisions are presented in
Figures 7 and 8 respectively. All the root-mean-squares, maximum
momentum components, and $p_T$ intervals are in the units of
GeV/$c$.
{%
\begin{center}
\begin{tabular}{ccccccc}
\hline\hline Type & $\sqrt{\overline{p_x^2}}$  & $\sqrt{\overline{p_y^2}}$  & $\sqrt{\overline{p_z^2}}$  & $|p_x|_{\max}$  & $|p_y|_{\max}$  & $|p_z|_{\max}$  \\
\hline
$pp$ & & & & & & \\
$1.7<p_{T}<2.0$   & $1.284\pm0.015$  & $1.322\pm0.014$  & $178.6\pm13.8$ & $1.991$  & $1.996$  & $1.314\times10^{3}$ \\
$6.7<p_{T}<7.7$   & $5.116\pm0.055$  & $4.970\pm0.057$  & $253.8\pm13.9$ & $7.679$  & $7.685$  & $1.369\times10^{3}$ \\
$19.9<p_{T}<22.8$ & $15.187\pm0.168$ & $14.780\pm0.173$ & $352.8\pm14.2$ & $22.730$ & $22.702$ & $1.376\times10^{3}$ \\
$59.8<p_{T}<94.8$ & $50.306\pm0.620$ & $49.707\pm0.612$ & $443.5\pm13.5$ & $93.992$ & $93.443$ & $1.361\times10^{3}$ \\
\hline
PbPb, 60--80\% & & & & & & \\
$1.7<p_{T}<2.0$   & $1.298\pm0.015$  & $1.304\pm0.015$  & $191.6\pm13.6$ & $1.998$  & $1.993$  & $1.354\times10^{3}$ \\
$6.7<p_{T}<7.7$   & $5.135\pm0.056$  & $4.946\pm0.058$  & $272.4\pm14.4$ & $7.655$  & $7.685$  & $1.367\times10^{3}$ \\
$19.9<p_{T}<22.8$ & $15.314\pm0.164$ & $14.654\pm0.171$ & $369.8\pm13.6$ & $22.730$ & $22.787$ & $1.343\times10^{3}$ \\
$59.8<p_{T}<94.8$ & $50.612\pm0.621$ & $50.001\pm0.611$ & $467.6\pm14.3$ & $94.461$ & $94.291$ & $1.371\times10^{3}$ \\
\hline
PbPb, 50--60\% & & & & & & \\
$1.7<p_{T}<2.0$   & $1.302\pm0.015$  & $1.298\pm0.015$  & $196.9\pm14.9$ & $1.995$  & $1.982$  & $1.374\times10^{3}$ \\
$6.7<p_{T}<7.7$   & $5.135\pm0.056$  & $4.956\pm0.058$  & $275.0\pm14.3$ & $7.655$  & $7.685$  & $1.368\times10^{3}$ \\
$19.9<p_{T}<22.8$ & $15.204\pm0.165$ & $14.766\pm0.170$ & $372.1\pm13.7$ & $22.730$ & $22.787$ & $1.361\times10^{3}$ \\
$59.8<p_{T}<94.8$ & $50.042\pm0.615$ & $50.514\pm0.614$ & $483.4\pm14.1$ & $94.461$ & $94.291$ & $1.373\times10^{3}$ \\
\hline
PbPb, 30--40\% & & & & & & \\
$1.7<p_{T}<2.0$   & $1.268\pm0.015$  & $1.335\pm0.014$  & $179.3\pm14.1$ & $1.995$  & $1.996$  & $1.346\times10^{3}$ \\
$6.7<p_{T}<7.7$   & $5.121\pm0.055$  & $4.968\pm0.057$  & $242.0\pm13.7$ & $7.655$  & $7.685$  & $1.360\times10^{3}$ \\
$19.9<p_{T}<22.8$ & $15.292\pm0.167$ & $14.715\pm0.174$ & $347.4\pm14.1$ & $22.730$ & $22.702$ & $1.375\times10^{3}$ \\
$59.8<p_{T}<94.8$ & $50.768\pm0.621$ & $49.803\pm0.612$ & $436.0\pm13.2$ & $94.062$ & $93.583$ & $1.369\times10^{3}$ \\
\hline
PbPb, 10--20\% & & & & & & \\
$1.7<p_{T}<2.0$   & $1.268\pm0.015$  & $1.335\pm0.014$  & $179.3\pm14.1$ & $1.995$  & $1.996$  & $1.346\times10^{3}$ \\
$6.7<p_{T}<7.7$   & $5.121\pm0.055$  & $4.968\pm0.057$  & $242.0\pm13.7$ & $7.655$  & $7.685$  & $1.360\times10^{3}$ \\
$19.9<p_{T}<22.8$ & $15.294\pm0.167$ & $14.717\pm0.174$ & $347.5\pm14.1$ & $22.730$ & $22.702$ & $1.375\times10^{3}$ \\
$59.8<p_{T}<94.8$ & $50.848\pm0.623$ & $49.886\pm0.613$ & $436.8\pm13.3$ & $94.062$ & $93.583$ & $1.369\times10^{3}$ \\
\hline
PbPb, 0--5\% & & & & & & \\
$1.7<p_{T}<2.0$   & $1.288\pm0.015$  & $1.310\pm0.014$  & $165.5\pm14.3$ & $1.995$  & $1.995$  & $1.333\times10^{3}$ \\
$6.7<p_{T}<7.7$   & $5.135\pm0.055$  & $4.955\pm0.057$  & $218.6\pm12.9$ & $7.655$  & $7.685$  & $1.372\times10^{3}$ \\
$19.9<p_{T}<22.8$ & $15.382\pm0.165$ & $14.633\pm0.173$ & $343.9\pm14.5$ & $22.581$ & $22.702$ & $1.376\times10^{3}$ \\
$59.8<p_{T}<94.8$ & $51.362\pm0.617$ & $49.605\pm0.620$ & $434.5\pm13.2$ & $94.097$ & $93.653$ & $1.368\times10^{3}$ \\
\hline
\end{tabular}
\end{center}
}} }

\vskip1.0cm {\scriptsize {Table 6. Percentages of particle numbers
in the interval $-50<p_z<50$ GeV/$c$ over whole $p_z$ range
corresponding to the scatter plots in different $p_T$ intervals in
different types of collisions, where the corresponding scatter
plots in $pp$ and 0--5\% Pb-Pb collisions are presented in Figures
7 and 8 respectively, and ``PL" denotes the ``power-law". All the
$p_T$ intervals are in the units of GeV/$c$.
{%
\begin{center}
\begin{tabular}{ccccccccc}
\hline\hline
& \multicolumn{2}{c}{$1.7<p_T<2.0$} & \multicolumn{2}{c}{$6.7<p_T<7.7$} & \multicolumn{2}{c}{$19.9<p_T<22.8$} & \multicolumn{2}{c}{$59.8<p_T<94.8$} \\
\cline{2-3} \cline{4-5} \cline{6-7} \cline{8-9}
Type & Inverse PL & Erlang & Inverse PL & Erlang & Inverse PL & Erlang & Inverse PL & Erlang \\
\hline
$pp$           & $78\%$ & $78\%$ & $61\%$ & $62\%$ & $43\%$ & $-$ & $21\%$ & $-$ \\
PbPb, 60--80\% & $75\%$ & $75\%$ & $58\%$ & $56\%$ & $41\%$ & $-$ & $20\%$ & $-$ \\
PbPb, 50--60\% & $75\%$ & $75\%$ & $58\%$ & $57\%$ & $41\%$ & $-$ & $20\%$ & $-$ \\
PbPb, 30--40\% & $80\%$ & $79\%$ & $62\%$ & $62\%$ & $43\%$ & $-$ & $20\%$ & $-$ \\
PbPb, 10--20\% & $80\%$ & $79\%$ & $62\%$ & $62\%$ & $43\%$ & $-$ & $21\%$ & $-$ \\
PbPb, 0--5\%   & $81\%$ & $81\%$ & $63\%$ & $62\%$ & $44\%$ & $-$ & $22\%$ & $-$ \\
\hline
\end{tabular}
\end{center}
}} } \vskip1.0cm

\begin{figure}
\hskip-1.0cm \begin{center}
\includegraphics[width=16.0cm]{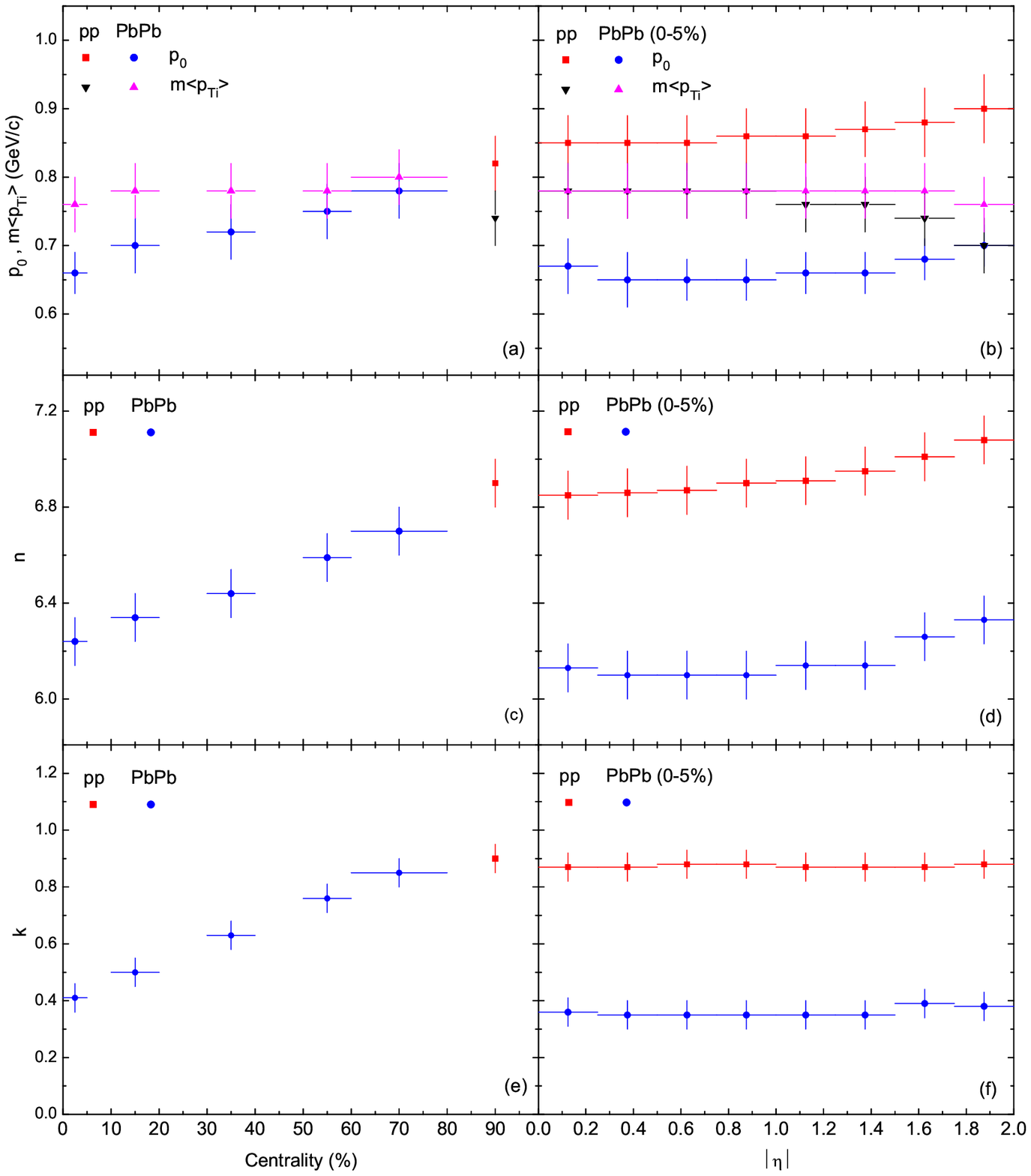}
\end{center}
\vskip.0cm Fig. 9. Left: Dependences of (a) $p_0$ and $m\langle
p_{Ti}\rangle$, (c) $n$, and (e) $k$ on centrality in $pp$ and
Pb-Pb collisions at 2.76 TeV, where the results corresponding to
$pp$ collisions are listed in 90\% centrality for comparisons.
Right: Dependences of (b) $p_0$ and $m\langle p_{Ti}\rangle$, (d)
$n$, and (f) $k$ on $|\eta|$ in $pp$ and 0--5\% Pb-Pb collisions.
Different symbols represent different collisions and quantities
shown in the panels, where the idiographic values are taken from
Table 1.
\end{figure}

\begin{figure}
\hskip-1.0cm \begin{center}
\includegraphics[width=10.0cm]{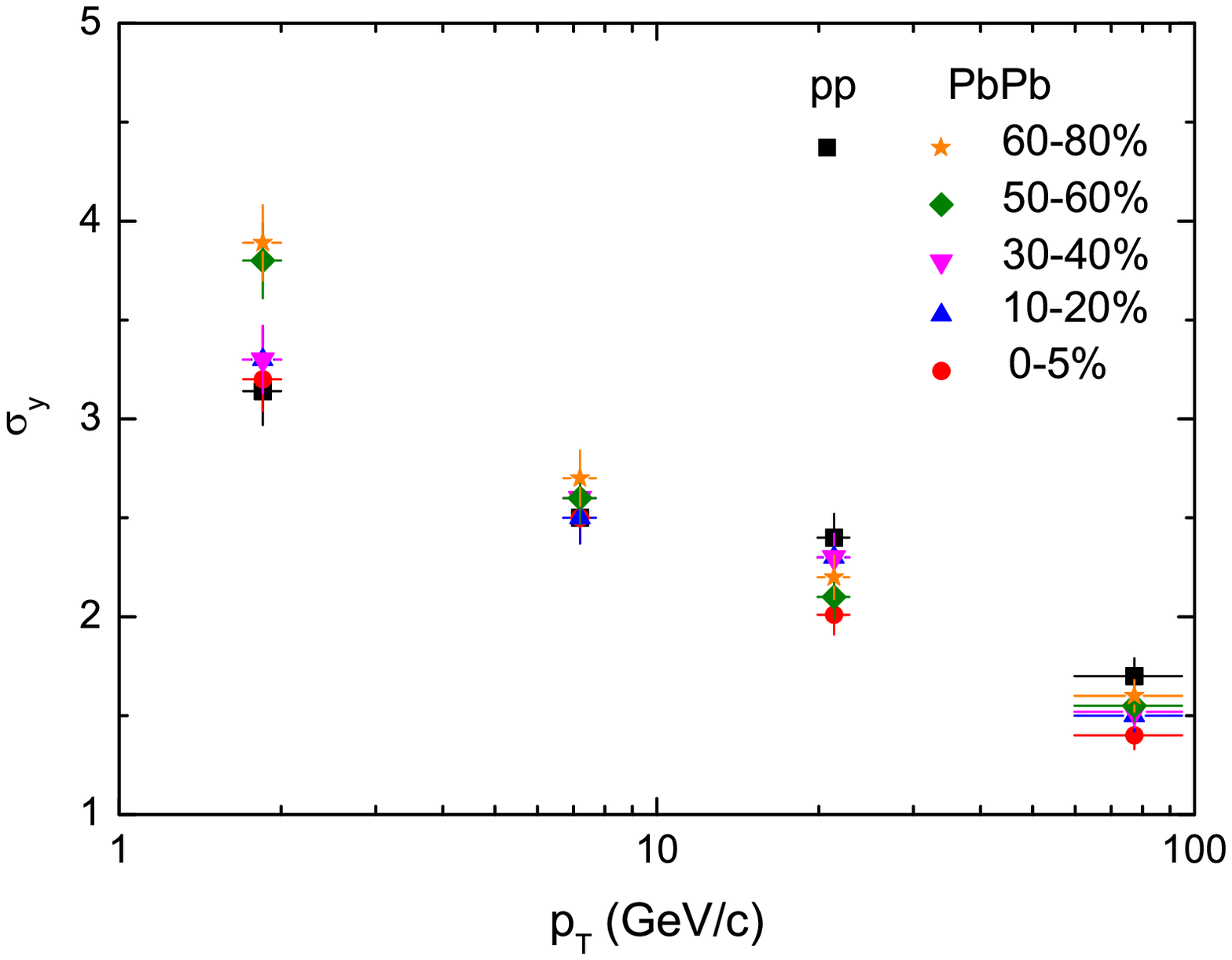}
\end{center}
\vskip.0cm Fig. 10. Dependence of $\sigma_y$ on $p_T$ in $pp$ and
Pb-Pb collisions at 2.76 TeV. Different symbols represent
different collisions shown in the panel, where the idiographic
values are taken from Table 2.
\end{figure}

By using the same method as that for Figure 7, we can obtain the
similar results in Pb-Pb collisions with centrality intervals
60--80\%, 50--60\%, 30--40\%, 10--20\%, and 0--5\%, respectively.
The values of root-mean squares of momentum components and the
maximum momentum components are listed in Table 5, and the
relative yields of particle numbers appearing in different $p_T$
intervals are listed in Table 4. As another example, to reduce
also the size of the paper file, only the scatter plots in the
three-dimensional momentum space in 0--5\% Pb-Pb collisions are
given in Figure 8. Some conclusions obtained from Figure 7 can be
obtained from Figure 8 and Tables 4 and 5. In addition, we see
intuitively the density change of particle numbers in the
three-dimensional momentum space in different $p_T$ intervals at
the kinetic freeze-out of the interacting system formed in Pb-Pb
collisions with different centrality intervals.

From Figures 5--8 and Tables 2, 3, and 5, one can see that the
hard scattering process that is described by the inverse power-law
corresponds to a wide $|p_i|$ ($i=x$, $y$, and $z$) range, large
$\sqrt{\overline{p_i^2}}$, wide $E$ range, wide $|\beta_{x,y}|$
range, and large $\sqrt{\overline{\beta_{x,y}^2}}$ due to large
momentum transfer and violent collisions between valence quarks.
At the same time, because $\beta_z$ is determined by $p_z/E$, we
are not sure to obtain a wide $|\beta_z|$ range and large
$\sqrt{\overline{\beta_z^2}}$ in the hard scattering process.
Instead, a narrow $|\beta_z|$ range and small
$\sqrt{\overline{\beta_z^2}}$ can be generally obtained, which
results in a narrow $|y|$ range and small $\sigma_y$. The
situation of the soft excitation process that is described by the
Erlang distribution is opposite. That is, in the soft process, we
can obtain a narrow $|p_i|$ range, small
$\sqrt{\overline{p_i^2}}$, narrow $E$ range, narrow
$|\beta_{x,y}|$ range, and small $\sqrt{\overline{\beta_{x,y}^2}}$
due to small momentum transfer and non-violent collisions between
gluons and/or sea quarks. Similarly, a wide $|\beta_z|$ range,
wide $|y|$ range, and large $\sigma_y$ can be obtained, too. These
differences between the two processes can be partly reflected in
the scatter plots.

As an example for comparison, in the three-dimensional velocity
space, the event patterns (particle scatter plots) presented in
the present work are obviously different from our recent work [41]
due to different types of particles being studied. Charged
particle scatter plots show that the root-mean-square velocities
form an ellipsoid surface with the major axis along the beam
direction, and the maximum velocities form a spherical surface.
Both the $Z$ boson and quarkonium state scatter plots show that
the root-mean square velocities form a rough cylinder surface
along the beam direction and the maximum velocities form a fat
cylinder surface which has the length being 1.2--2.2 times of
diameter, due to their productions being at the initial stage of
collisions. Contrastively, charged particles which are mainly
pions produce mostly at the intermediate stage of collisions and
suffer particularly the processes of thermalization and expansion
of the interacting system. Generally, different scatter plots
reflect different production stages of different types of
particles. The present work shows similar result to our another
recent work [40] which studies the scatter plots of net-baryons
which are a part of charged particles and suffer the
thermalization and expansion of the system.

We would like to point out that a correct description of the low
$p_T$ part using the Erlang distribution should seemingly include
the effect of transverse flow. In fact, the parameters extracted
from the $p_T$ spectra contain naturally the contribution of flow
effect, though the flow effect and thermal motion are tangled each
other. Only in the case of extracting the kinetic freeze-out
temperature, the flow effect should be eliminated [44--46]. The
present work focuses mainly on quantities at the kinetic
freeze-out, but it does not include the kinetic freeze-out
temperature which needs more spectra of identified particles. Both
the contributions of flow effect and thermal motion are contained
in the extraction of parameters, though the two contributions are
not distinguished. To severally understand the contributions of
flow effect and other non-flow effects such as thermal motion,
more than one disentangling methods are used in literature [47,
48]. In our very recent works [44--46], an alternative method is
used to disentangle the two effects.

In the alternative method, the collective expansion of source and
random thermal motion of particles are naturally disentangled to
be the mean transverse flow velocity and kinetic freeze-out
temperature, respectively. The mean transverse flow velocity can
be extracted from the slope of the mean transverse momentum
($\langle p_T \rangle$) curve as a function of mean moving mass
($\overline {m}$) when plotting $\langle p_T \rangle$ versus
$\overline {m}$, and the kinetic freeze-out temperature can be
extracted from the intercept of the effective temperature ($T$)
curve as a function of $m_0$ when plotting $T$ versus $m_0$ [46],
where $T$ is usually regarded as the inverse slopes of $p_T$
spectra and $m$ is also used to denote the moving mass. To use
this alternative method, we need at least $p_T$ spectra of pions,
kaons, and protons in the same experimental condition. As
statistical results, the flow effect and thermal motion can be
disentangled in principle. In the case of disentangling the two
effects, their several contributions can be obtained. It is
regretful that the charged particle spectra discussed in the
present work are not suitable to use the alternative method.

\begin{figure}
\hskip-1.0cm \begin{center}
\includegraphics[width=16.0cm]{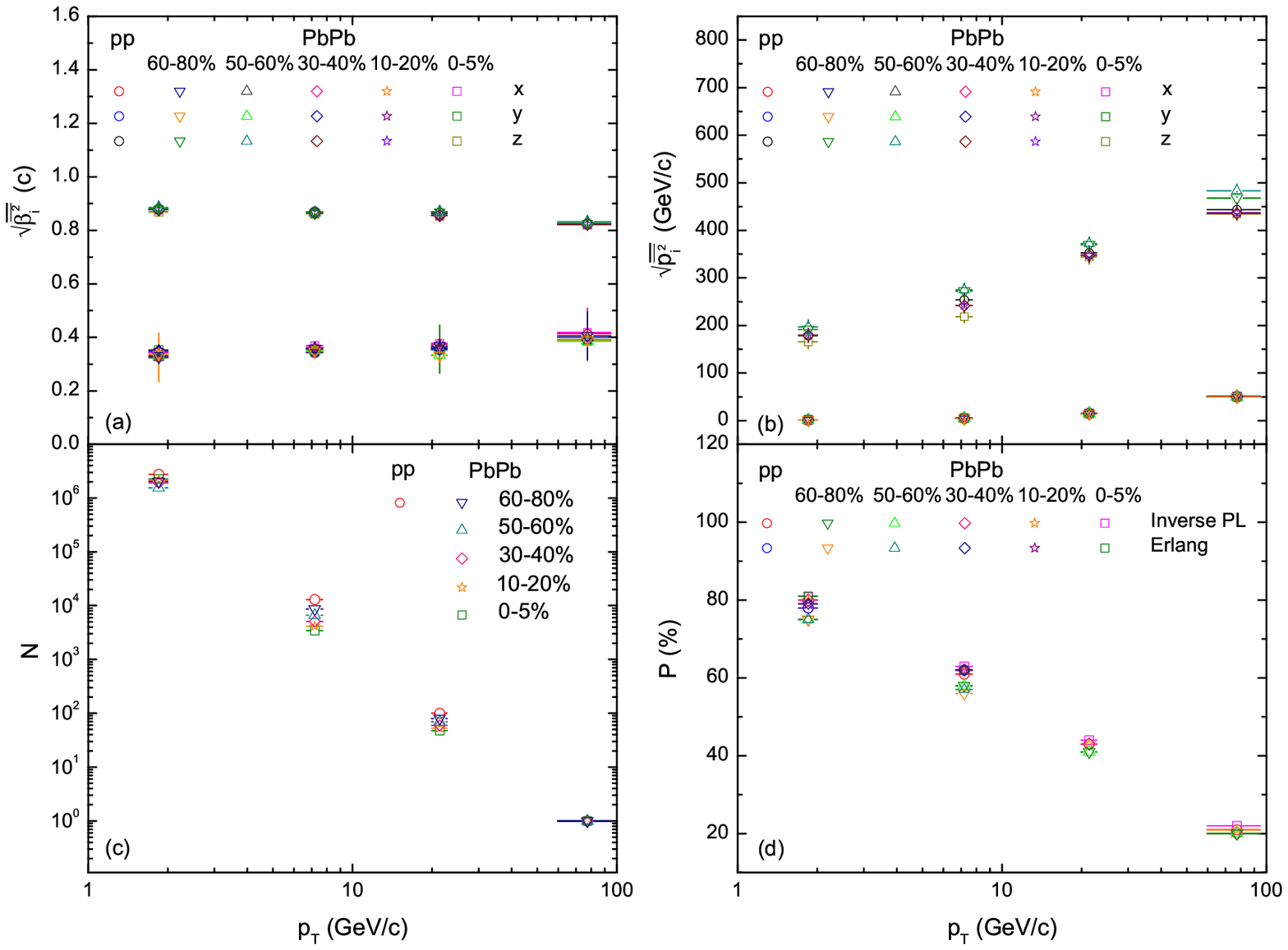}
\end{center}
\vskip.0cm Fig. 11. Dependences of (a)
$\sqrt{\overline{\beta^2_i}}$ ($i=x$, $y$, and $z$), (b)
$\sqrt{\overline{p^2_i}}$, (c) relative particle number $N$, and
(d) percentage $P$ in $-50<p_z<50$ GeV/$c$ over whole $p_z$ range
on $p_T$ in $pp$ and Pb-Pb collisions at 2.76 TeV. Different
symbols represent different collisions and quantities shown in the
panels, where the idiographic values are taken from Tables 3, 5,
4, and 6, respectively.
\end{figure}

To study in detail the tendencies of free parameters, the left
panel in Figure 9 shows the dependences of (a) $p_0$ and $m\langle
p_{Ti}\rangle$, (c) $n$, and (e) $k$ on centrality in $pp$ and
Pb-Pb collisions at 2.76 TeV, where the results corresponding to
$pp$ collisions are listed in 90\% centrality for comparisons. The
right panel in Figure 9 shows the dependences of (b) $p_0$ and
$m\langle p_{Ti}\rangle$, (d) $n$, and (f) $k$ on $|\eta|$ in $pp$
and 0--5\% Pb-Pb collisions. Figure 10 shows the dependence of
$\sigma_y$ on $p_T$ in $pp$ and Pb-Pb collisions. In Figures 9 and
10, different symbols represent different collisions and
quantities shown in the panels, where the idiographic values are
taken from Tables 1 and 2. One can see that $p_0$, $n$, and $k$
increase with the decrease of centrality. $m\langle p_{Ti}\rangle$
does not show an obvious tendency with the centrality. In $pp$
collisions, $p_0$ increases slightly and $m\langle p_{Ti}\rangle$
decreases slightly with the increase of $|\eta|$. In 0--5\% Pb-Pb
collisions, $p_0$ and $m\langle p_{Ti}\rangle$ do not show an
obvious tendency with the increase of $|\eta|$. In both $pp$ and
0--5\% Pb-Pb collisions, $n$ increases slightly and $k$ does not
change with the increase of $|\eta|$. The tendency that $\sigma_y$
decreases with the increases of $p_T$ does not obviously depend on
system size and collision centrality.

We can explain the characteristics of some parameters. For hard
process, the values of $p_0$, $n$, and $k$ in $pp$ or peripheral
Pb-Pb collisions are greater than those in central Pb-Pb
collisions due to more successive nucleon-nucleon collisions in
the latter one. Although the successive nucleon-nucleon collisions
can produce more particles, the violent degree in head-on
scattering between valence quarks is reduced, which renders small
values of parameters. For soft process, the values of $m\langle
p_{Ti}\rangle$ in collisions for different sizes and centralities
are close to each other due to the similar excitation degree
between gluons and/or sea quarks and the similar flow effect. On
the dependences of $p_0$ and $m\langle p_{Ti}\rangle$ on $|\eta|$
in $pp$ collisions, different tendencies appear due to different
participant partons. In 0--5\% Pb-Pb collisions, the different
tendencies are reduced due to more intranuclear cascade
collisions. As for $n$ and $k$, the same tendency appears for both
$pp$ and 0--5\% Pb-Pb collisions, though the values for $pp$
collisions are greater than those for 0--5\% Pb-Pb collisions due
to absent intranuclear process in $pp$ collisions. The decrescent
tendency of $\sigma_y$ with increasing $p_T$ is a natural result,
where small $p_T$ corresponds to small/large angle and large
$|\eta|$, and large $p_T$ corresponds to middle angle and small
$|\eta|$. For the production of charged particles, the dependence
of free parameters on collision centrality is a reflection of cold
nuclear effect which results from different numbers of
multi-scattering in cold nuclei or spectators with different
sizes. The larger the spectator is, the more the parameters
depend.

According to Tables 3--6, we obtain the dependences of
$\sqrt{\overline{\beta^2_i}}$, $\sqrt{\overline{p^2_i}}$, relative
particle number $N$, and percentage $P$ in $-50<p_z<50$ GeV/$c$
over whole $p_z$ range, on $p_T$ in Figure 11, where different
symbols represent different collisions and quantities shown in the
panels. One can see that $\sqrt{\overline{\beta^2_x}}$ and
$\sqrt{\overline{\beta^2_y}}$ increase slightly with the increase
of $p_T$, and they are almost the same for different sizes and
centralities of collisions. $\sqrt{\overline{\beta^2_z}}$
decreases slightly with the increase of $p_T$, and it is almost
the same for different sizes and centralities. The situations for
$\sqrt{\overline{p^2_x}}$ and $\sqrt{\overline{p^2_y}}$ are the
same as those for $\sqrt{\overline{\beta^2_x}}$ and
$\sqrt{\overline{\beta^2_y}}$. $\sqrt{\overline{p^2_z}}$ increases
obviously with the increase of $p_T$, where the values for
different sizes and centralities are distinguishable, though the
characteristic of distinction is not obvious. $N$ decreases
quickly with the increase of $p_T$, and in some cases the values
are almost the same for different sizes and centralities. $P$ also
decreases with the increase of $p_T$ for both inverse power-law
and Erlang distribution, and in some cases the values are very
close to each other for different sizes, centralities, and
functions. The characteristics presented in Figure 11 are
determined by the parameters presented in Tables 1 and 2 (or
Figures 9 and 10).

From the above discussions, although one can see a very complete
analysis of $p_T$ and $y$ spectra of charged particles produced in
$pp$ and Pb-Pb collisions at the LHC, it seems that the model used
for this analysis does not have a strong theoretical basis. In
fact, we have used a hybrid model in which each part has its
reason. For the $p_T$ spectra, it considers the superposition of a
polynomial inverse power-law which is suggested by the QCD
calculus [18--20] and an exponential-like Erlang distribution
which is resulted from the multisource thermal model [17]. For the
$y$ spectra, it considers a Gaussian function which is resulted
from the Landau hydrodynamic model [21--24]. Both the functions
for $p_T$ and $y$ spectra look very simple, useful, and efficient.
In particular, the two-component function for $p_T$ spectra
contains both the contributions of hard scattering and soft
excitation processes which correspond to violent collisions
between valence quarks and non-violent collisions between gluons
and/or sea quarks respectively. At the same time, the hard and
soft processes result from large and small momentum transfers, and
contribute in wide and narrow $p_T$ regions, respectively. At the
considered energy, in most cases, the hard process has a large
contribution, which is different from that at low energy.
\\

{\section{Conclusions}}

We summarize here our main observations and conclusions.

(a) The transverse momentum and pseudorapidity spectra of charged
particles produced in $pp$ collisions at $\sqrt{s}=2.76$ TeV and
in Pb-Pb collisions with different centrality intervals at
$\sqrt{s_{NN}}=2.76$ TeV at the LHC are described by the hybrid
model, in which the two-component $p_T$ distribution (which
contains the inverse power-law and the Erlang distribution) and
the Gaussian $y$ distribution are used. The inverse power-law is
based on the QCD calculus, the Erlang distribution is resulted
from the multisource thermal model, and the Gaussian $y$
distribution is resulted from the Landau hydrodynamic model. The
modelling results are in agreement with the experimental data of
the ATLAS Collaboration.

(b) In the hybrid model, both the functions for $p_T$ and $y$
spectra look very simple, useful, and efficient. In particular,
the two-component function for $p_T$ spectra contains both the
contributions of hard scattering and soft excitation processes
which correspond to violent collisions between valence quarks and
non-violent collisions between gluons and/or sea quarks
respectively. Not only for the hard scattering process but also
for the soft excitation process, the numbers of participant
partons are two, in which one is the projectile parton and the
other one is the target parton. The hard scattering process
contributes a wide $p_T$ range, and the soft excitation process
contributes a narrow $p_T$ range. In the considered collisions at
the LHC, the hard process has a large contribution in most cases.

(c) The parameters $p_0$, $n$, and $k$ increase with the decrease
of centrality. $m\langle p_{Ti}\rangle$ does not show an obvious
tendency with the centrality. For hard process, the values of
$p_0$, $n$, and $k$ in $pp$ or peripheral Pb-Pb collisions are
greater than those in central Pb-Pb collisions due to more
successive nucleon-nucleon collisions in the latter one. For soft
process, the values of $m\langle p_{Ti}\rangle$ in collisions for
different sizes and centralities are close to each other due to
the similar excitation degree between gluons and/or sea quarks and
the similar flow effect. The tendency that $\sigma_y$ decreases
with the increase of $p_T$ is a natural result due to small $p_T$
corresponding to small/large angle and large $|\eta|$, and large
$p_T$ corresponding to middle angle and small $|\eta|$.

(d) In $pp$ collisions, $p_0$ slightly increases and $m\langle
p_{Ti}\rangle$ slightly decreases with the increase of $|\eta|$.
In 0--5\% Pb-Pb collisions, $p_0$ and $m\langle p_{Ti}\rangle$ do
not show an obvious tendency with the increase of $|\eta|$. On the
dependence of $p_0$ and $m\langle p_{Ti}\rangle$ on $|\eta|$ in
$pp$ collisions, different tendencies appear due to different
participant partons (valence quarks versus gluons and/or sea
quarks). In 0--5\% Pb-Pb collisions, the different tendencies in
the dependence of $p_0$ and $m\langle p_{Ti}\rangle$ on $|\eta|$
are reduced due to more intranuclear collisions. In both $pp$ and
0--5\% Pb-Pb collisions, $n$ slightly increases and $k$ does not
change with the increase of $|\eta|$. The values of $n$ and $k$
for $pp$ collisions are greater than those for 0--5\% Pb-Pb
collisions due to absent intranuclear process in $pp$ collisions.

(e) Based on the parameter values extracted from $p_T$ and $\eta$
spectra, the event patterns (particle scatter plots) in the
three-dimensional velocity and momentum spaces are obtained. In
particular, $\sqrt{\overline{\beta_x^2}} \approx
\sqrt{\overline{\beta_y^2}} \ll \sqrt{\overline{\beta_z^2}}$,
$|\beta_x|_{\max} \approx |\beta_y|_{\max} \approx
|\beta_z|_{\max} \approx 1$, $\sqrt{\overline{p_x^2}} \approx
\sqrt{\overline{p_y^2}} \ll \sqrt{\overline{p_z^2}}$, and
$|p_x|_{\max} \approx |p_y|_{\max} \ll |p_z|_{\max}$. The
root-mean-square velocities form an ellipsoid surface with the
major axis along the beam direction, and the maximum velocities
form a spherical surface. Viewing the wide $p_z$ range, most
particles constitute a circle-like region surrounded by a few
particles; and viewing the narrow $p_z$ range, particles
constitute a cylinder surface surrounded by a few particles. The
severally relative sizes of $\sqrt{\overline{\beta^2_i}}$,
$|\beta_i|_{\max}$, $\sqrt{\overline{p^2_i}}$, and $|p_i|_{\max}$
for different $i$, as well as the related characteristics are
determined by the extracted parameters and isotropic assumption in
transverse plane.

(f) Further, based on the parameters extracted above,
$\sqrt{\overline{\beta^2_x}}$ and $\sqrt{\overline{\beta^2_y}}$
slightly increase with the increase of $p_T$, and they are almost
the same for different sizes and centralities of collisions. The
situations for $\sqrt{\overline{p^2_x}}$ and
$\sqrt{\overline{p^2_y}}$ are the same as those for
$\sqrt{\overline{\beta^2_x}}$ and $\sqrt{\overline{\beta^2_y}}$.
$\sqrt{\overline{\beta^2_z}}$ slightly decreases with the increase
of $p_T$, and it is almost the same for different sizes and
centralities. $\sqrt{\overline{p^2_z}}$ obviously increases with
the increase of $p_T$, where the values for different sizes and
centralities are distinguishable, though the characteristic of
distinction is not obvious. Naturally, the characteristics of
$\sqrt{\overline{\beta^2_i}}$ and $\sqrt{\overline{p^2_i}}$ are
determined by the extracted parameters, which are determined by
many factors such as the number of intranuclear cascade
collisions, type of participant partons, dependence of collision
centrality (number of multi-scattering in cold nucleus or
spectator), and others.

(g) The hard scattering process corresponds to a wide $|p_i|$
range, large $\sqrt{\overline{p_i^2}}$, wide $E$ range, wide
$|\beta_{x,y}|$ range, large $\sqrt{\overline{\beta_{x,y}^2}}$,
narrow $|\beta_z|$ range, small $\sqrt{\overline{\beta_z^2}}$,
narrow $|y|$ range, and small $\sigma_y$. The situation of the
soft excitation process is opposite. These differences between the
two processes can be partly reflected in the scatter plots in
three-dimensional velocity and momentum spaces where the hard
process corresponds to wider range. The reason that causes these
differences is different interacting mechanisms. Generally, the
hard scattering process is produced due to violent collisions
between valence quarks where large momentum transfer occurs. The
soft excitation process is produced due to non-violent collisions
between gluons and/or sea quarks where small momentum transfer
occurs.

(h) Different types of particles correspond to different scatter
plots due to different production stages. The scatter plots of
charged particles are different from those of $Z$ bosons and
quarkonium states discussed in our recent work [41] in which the
root-mean square velocities show a rough cylinder surface and the
maximum velocities form a fat cylinder surface in the
three-dimensional velocity space, due to the productions of $Z$
bosons and quarkonium states being at the initial stage of
collisions, while charged particles which are mainly pions produce
mostly at the intermediate stage of collisions and suffer
particularly the processes of thermalization and expansion of the
interacting system. Combining with our recent works [40, 41], we
have provided a reference in methodology which displays event
patterns (particle scatter plots) for different particles in
three-dimensional velocity and momentum spaces or other available
spaces based on the transverse momentum and pseudorapidity or
rapidity spectra of considered particles.
\\

{\bf Acknowledgments}

This work was supported by the National Natural Science Foundation
of China under Grant No. 11575103.

\end{document}